\documentclass[longauth]{aa}

\usepackage{graphicx}
\usepackage{xcolor}
\usepackage{txfonts}
\usepackage{euclid}
\usepackage{lastpage}
\usepackage{multirow}
\usepackage{tabularx}
\usepackage{amsmath}
 \usepackage{booktabs}
\usepackage{svg}
\graphicspath{{./}{figures/}}
\usepackage[pdfencoding=auto,psdextra]{hyperref}
\hypersetup{
    colorlinks=true,
    linkcolor=blue,
    filecolor=magenta,      
    urlcolor=blue,
    citecolor=blue
}
\usepackage[nameinlink,capitalise]{cleveref}
\crefname{section}{Sect.}{Sects.}
\Crefname{section}{Section}{Sections}
\crefname{figure}{Fig.}{Figs.}
\Crefname{figure}{Figure}{Figures}
\crefname{equation}{Eq.}{Eqs.}
\Crefname{equation}{Equation}{Equations}
\crefname{table}{Table}{Tables}
\crefname{appendix}{Appendix}{Appendices}

\newcommand{\lsim}{\raisebox{-0.13cm}{~\shortstack{$<$ \\[-0.07cm] $\sim$}}~}
\newcommand{\gsim}{\raisebox{-0.13cm}{~\shortstack{$>$ \\[-0.07cm] $\sim$}}~}

\usepackage[utf8]{inputenc}

\usepackage[switch, modulo]{lineno}

\usepackage{euclid}

\begin{document} 


\title{\Euclid: Scaled-up little red dots and other sources with v-shaped spectral energy distributions at $z>4$\thanks{This paper is published on behalf of the Euclid Consortium}}

\titlerunning{\Euclid: Scaled-up LRDs and other v-shaped SED sources at $z>4$}
\authorrunning{Tumborang et al.}

\newcommand{\orcid}[1]{} 


\author{A.~A.~Tumborang\orcid{0009-0008-3005-0435}\thanks{\email{tumborang@astro.rug.nl}}\inst{\ref{aff1}}
\and K.~I.~Caputi\orcid{0000-0001-8183-1460}\inst{\ref{aff1},\ref{aff2}}
\and P.~Rinaldi\orcid{0000-0002-5104-8245}\inst{\ref{aff3}}
\and L.~Bisigello\orcid{0000-0003-0492-4924}\inst{\ref{aff4}}
\and G.~Girardi\orcid{0009-0005-6156-4066}\inst{\ref{aff5},\ref{aff4}}
\and E.~Iani\orcid{0000-0001-8386-3546}\inst{\ref{aff6}}
\and R.~Bouwens\orcid{0000-0002-4989-2471}\inst{\ref{aff7}}
\and R.~Navarro-Carrera\orcid{0000-0001-6066-4624}\inst{\ref{aff1}}
\and G.~Desprez\orcid{0000-0001-8325-1742}\inst{\ref{aff1}}
\and R.~A.~Cooper\orcid{0009-0000-0413-5699}\inst{\ref{aff1}}
\and Y.~Fu\orcid{0000-0002-0759-0504}\inst{\ref{aff7},\ref{aff1}}
\and Y.~Toba\orcid{0000-0002-3531-7863}\inst{\ref{aff8},\ref{aff9}}
\and J.~Matthee\orcid{0000-0003-2871-127X}\inst{\ref{aff6}}
\and B.~Milvang-Jensen\orcid{0000-0002-2281-2785}\inst{\ref{aff2},\ref{aff10}}
\and P.~G.~Perez-Gonzalez\orcid{0000-0003-4528-5639}\inst{\ref{aff11}}
\and F.~Ricci\orcid{0000-0001-5742-5980}\inst{\ref{aff12},\ref{aff13}}
\and G.~Rodighiero\orcid{0000-0002-9415-2296}\inst{\ref{aff5},\ref{aff4}}
\and J.~Schaye\orcid{0000-0002-0668-5560}\inst{\ref{aff7}}
\and F.~Tarsitano\orcid{0000-0002-5919-0238}\inst{\ref{aff14},\ref{aff15},\ref{aff16}}
\and G.~Zamorani\orcid{0000-0002-2318-301X}\inst{\ref{aff17}}
\and M.~Baes\orcid{0000-0002-3930-2757}\inst{\ref{aff18}}
\and C.~M.~Gutierrez\orcid{0000-0001-7854-783X}\inst{\ref{aff20},\ref{aff21}}
\and H.~Hoekstra\orcid{0000-0002-0641-3231}\inst{\ref{aff7}}
\and K.~Jahnke\orcid{0000-0003-3804-2137}\inst{\ref{aff22}}
\and D.~Scott\orcid{0000-0002-6878-9840}\inst{\ref{aff23}}
\and D.~Stern\orcid{0000-0003-2686-9241}\inst{\ref{aff24}}
\and B.~Altieri\orcid{0000-0003-3936-0284}\inst{\ref{aff25}}
\and S.~Andreon\orcid{0000-0002-2041-8784}\inst{\ref{aff26}}
\and N.~Auricchio\orcid{0000-0003-4444-8651}\inst{\ref{aff17}}
\and C.~Baccigalupi\orcid{0000-0002-8211-1630}\inst{\ref{aff27},\ref{aff28},\ref{aff29},\ref{aff30}}
\and M.~Baldi\orcid{0000-0003-4145-1943}\inst{\ref{aff31},\ref{aff17},\ref{aff32}}
\and A.~Balestra\orcid{0000-0002-6967-261X}\inst{\ref{aff4}}
\and S.~Bardelli\orcid{0000-0002-8900-0298}\inst{\ref{aff17}}
\and P.~Battaglia\orcid{0000-0002-7337-5909}\inst{\ref{aff17}}
\and A.~Biviano\orcid{0000-0002-0857-0732}\inst{\ref{aff28},\ref{aff27}}
\and E.~Branchini\orcid{0000-0002-0808-6908}\inst{\ref{aff33},\ref{aff34},\ref{aff26}}
\and M.~Brescia\orcid{0000-0001-9506-5680}\inst{\ref{aff35},\ref{aff36}}
\and S.~Camera\orcid{0000-0003-3399-3574}\inst{\ref{aff37},\ref{aff38},\ref{aff39}}
\and V.~Capobianco\orcid{0000-0002-3309-7692}\inst{\ref{aff39}}
\and C.~Carbone\orcid{0000-0003-0125-3563}\inst{\ref{aff40}}
\and J.~Carretero\orcid{0000-0002-3130-0204}\inst{\ref{aff41},\ref{aff42}}
\and S.~Casas\orcid{0000-0002-4751-5138}\inst{\ref{aff43},\ref{aff44}}
\and M.~Castellano\orcid{0000-0001-9875-8263}\inst{\ref{aff13}}
\and G.~Castignani\orcid{0000-0001-6831-0687}\inst{\ref{aff17}}
\and S.~Cavuoti\orcid{0000-0002-3787-4196}\inst{\ref{aff36},\ref{aff45}}
\and K.~C.~Chambers\orcid{0000-0001-6965-7789}\inst{\ref{aff46}}
\and A.~Cimatti\inst{\ref{aff47}}
\and C.~Colodro-Conde\inst{\ref{aff20}}
\and G.~Congedo\orcid{0000-0003-2508-0046}\inst{\ref{aff48}}
\and C.~J.~Conselice\orcid{0000-0003-1949-7638}\inst{\ref{aff49}}
\and L.~Conversi\orcid{0000-0002-6710-8476}\inst{\ref{aff50},\ref{aff25}}
\and Y.~Copin\orcid{0000-0002-5317-7518}\inst{\ref{aff51}}
\and F.~Courbin\orcid{0000-0003-0758-6510}\inst{\ref{aff52},\ref{aff53},\ref{aff54}}
\and H.~M.~Courtois\orcid{0000-0003-0509-1776}\inst{\ref{aff55}}
\and M.~Cropper\orcid{0000-0003-4571-9468}\inst{\ref{aff56}}
\and J.-G.~Cuby\orcid{0000-0002-8767-1442}\inst{\ref{aff57},\ref{aff58}}
\and A.~Da~Silva\orcid{0000-0002-6385-1609}\inst{\ref{aff59},\ref{aff60}}
\and H.~Degaudenzi\orcid{0000-0002-5887-6799}\inst{\ref{aff16}}
\and G.~De~Lucia\orcid{0000-0002-6220-9104}\inst{\ref{aff28}}
\and H.~Dole\orcid{0000-0002-9767-3839}\inst{\ref{aff61}}
\and M.~Douspis\orcid{0000-0003-4203-3954}\inst{\ref{aff61}}
\and F.~Dubath\orcid{0000-0002-6533-2810}\inst{\ref{aff16}}
\and X.~Dupac\inst{\ref{aff25}}
\and M.~Farina\orcid{0000-0002-3089-7846}\inst{\ref{aff62}}
\and R.~Farinelli\inst{\ref{aff17}}
\and F.~Faustini\orcid{0000-0001-6274-5145}\inst{\ref{aff13},\ref{aff63}}
\and S.~Ferriol\inst{\ref{aff51}}
\and F.~Finelli\orcid{0000-0002-6694-3269}\inst{\ref{aff17},\ref{aff64}}
\and S.~Fotopoulou\orcid{0000-0002-9686-254X}\inst{\ref{aff65}}
\and N.~Fourmanoit\orcid{0009-0005-6816-6925}\inst{\ref{aff66}}
\and M.~Frailis\orcid{0000-0002-7400-2135}\inst{\ref{aff28}}
\and E.~Franceschi\orcid{0000-0002-0585-6591}\inst{\ref{aff17}}
\and M.~Fumana\orcid{0000-0001-6787-5950}\inst{\ref{aff40}}
\and S.~Galeotta\orcid{0000-0002-3748-5115}\inst{\ref{aff28}}
\and K.~George\orcid{0000-0002-1734-8455}\inst{\ref{aff67}}
\and B.~Gillis\orcid{0000-0002-4478-1270}\inst{\ref{aff48}}
\and C.~Giocoli\orcid{0000-0002-9590-7961}\inst{\ref{aff17},\ref{aff32}}
\and J.~Gracia-Carpio\orcid{0000-0003-4689-3134}\inst{\ref{aff68}}
\and A.~Grazian\orcid{0000-0002-5688-0663}\inst{\ref{aff4}}
\and F.~Grupp\inst{\ref{aff68},\ref{aff69}}
\and S.~V.~H.~Haugan\orcid{0000-0001-9648-7260}\inst{\ref{aff70}}
\and W.~Holmes\inst{\ref{aff24}}
\and I.~M.~Hook\orcid{0000-0002-2960-978X}\inst{\ref{aff71}}
\and F.~Hormuth\inst{\ref{aff72}}
\and A.~Hornstrup\orcid{0000-0002-3363-0936}\inst{\ref{aff73},\ref{aff74}}
\and M.~Jhabvala\inst{\ref{aff75}}
\and B.~Joachimi\orcid{0000-0001-7494-1303}\inst{\ref{aff76}}
\and S.~Kermiche\orcid{0000-0002-0302-5735}\inst{\ref{aff66}}
\and A.~Kiessling\orcid{0000-0002-2590-1273}\inst{\ref{aff24}}
\and B.~Kubik\orcid{0009-0006-5823-4880}\inst{\ref{aff51}}
\and M.~K\"ummel\orcid{0000-0003-2791-2117}\inst{\ref{aff69}}
\and M.~Kunz\orcid{0000-0002-3052-7394}\inst{\ref{aff77}}
\and H.~Kurki-Suonio\orcid{0000-0002-4618-3063}\inst{\ref{aff78},\ref{aff79}}
\and A.~M.~C.~Le~Brun\orcid{0000-0002-0936-4594}\inst{\ref{aff80}}
\and S.~Ligori\orcid{0000-0003-4172-4606}\inst{\ref{aff39}}
\and P.~B.~Lilje\orcid{0000-0003-4324-7794}\inst{\ref{aff70}}
\and V.~Lindholm\orcid{0000-0003-2317-5471}\inst{\ref{aff78},\ref{aff79}}
\and I.~Lloro\orcid{0000-0001-5966-1434}\inst{\ref{aff81}}
\and G.~Mainetti\orcid{0000-0003-2384-2377}\inst{\ref{aff82}}
\and E.~Maiorano\orcid{0000-0003-2593-4355}\inst{\ref{aff17}}
\and O.~Mansutti\orcid{0000-0001-5758-4658}\inst{\ref{aff28}}
\and S.~Marcin\inst{\ref{aff83}}
\and O.~Marggraf\orcid{0000-0001-7242-3852}\inst{\ref{aff84}}
\and M.~Martinelli\orcid{0000-0002-6943-7732}\inst{\ref{aff13},\ref{aff85}}
\and N.~Martinet\orcid{0000-0003-2786-7790}\inst{\ref{aff58}}
\and F.~Marulli\orcid{0000-0002-8850-0303}\inst{\ref{aff86},\ref{aff17},\ref{aff32}}
\and R.~J.~Massey\orcid{0000-0002-6085-3780}\inst{\ref{aff87}}
\and E.~Medinaceli\orcid{0000-0002-4040-7783}\inst{\ref{aff17}}
\and S.~Mei\orcid{0000-0002-2849-559X}\inst{\ref{aff88},\ref{aff89}}
\and M.~Meneghetti\orcid{0000-0003-1225-7084}\inst{\ref{aff17},\ref{aff32}}
\and E.~Merlin\orcid{0000-0001-6870-8900}\inst{\ref{aff13}}
\and G.~Meylan\inst{\ref{aff90}}
\and A.~Mora\orcid{0000-0002-1922-8529}\inst{\ref{aff91}}
\and M.~Moresco\orcid{0000-0002-7616-7136}\inst{\ref{aff86},\ref{aff17}}
\and L.~Moscardini\orcid{0000-0002-3473-6716}\inst{\ref{aff86},\ref{aff17},\ref{aff32}}
\and C.~Neissner\orcid{0000-0001-8524-4968}\inst{\ref{aff92},\ref{aff42}}
\and R.~C.~Nichol\orcid{0000-0003-0939-6518}\inst{\ref{aff93}}
\and S.-M.~Niemi\orcid{0009-0005-0247-0086}\inst{\ref{aff94}}
\and J.~W.~Nightingale\orcid{0000-0002-8987-7401}\inst{\ref{aff95}}
\and C.~Padilla\orcid{0000-0001-7951-0166}\inst{\ref{aff92}}
\and S.~Paltani\orcid{0000-0002-8108-9179}\inst{\ref{aff16}}
\and F.~Pasian\orcid{0000-0002-4869-3227}\inst{\ref{aff28}}
\and K.~Pedersen\inst{\ref{aff96}}
\and W.~J.~Percival\orcid{0000-0002-0644-5727}\inst{\ref{aff97},\ref{aff98},\ref{aff99}}
\and V.~Pettorino\orcid{0000-0002-4203-9320}\inst{\ref{aff94}}
\and S.~Pires\orcid{0000-0002-0249-2104}\inst{\ref{aff100}}
\and G.~Polenta\orcid{0000-0003-4067-9196}\inst{\ref{aff63}}
\and M.~Poncet\inst{\ref{aff101}}
\and L.~A.~Popa\inst{\ref{aff102}}
\and L.~Pozzetti\orcid{0000-0001-7085-0412}\inst{\ref{aff17}}
\and F.~Raison\orcid{0000-0002-7819-6918}\inst{\ref{aff68}}
\and A.~Renzi\orcid{0000-0001-9856-1970}\inst{\ref{aff5},\ref{aff103}}
\and J.~Rhodes\orcid{0000-0002-4485-8549}\inst{\ref{aff24}}
\and G.~Riccio\inst{\ref{aff36}}
\and E.~Romelli\orcid{0000-0003-3069-9222}\inst{\ref{aff28}}
\and M.~Roncarelli\orcid{0000-0001-9587-7822}\inst{\ref{aff17}}
\and B.~Rusholme\orcid{0000-0001-7648-4142}\inst{\ref{aff104}}
\and R.~Saglia\orcid{0000-0003-0378-7032}\inst{\ref{aff69},\ref{aff68}}
\and Z.~Sakr\orcid{0000-0002-4823-3757}\inst{\ref{aff105},\ref{aff106},\ref{aff107}}
\and D.~Sapone\orcid{0000-0001-7089-4503}\inst{\ref{aff108}}
\and M.~Schirmer\orcid{0000-0003-2568-9994}\inst{\ref{aff22}}
\and P.~Schneider\orcid{0000-0001-8561-2679}\inst{\ref{aff84}}
\and T.~Schrabback\orcid{0000-0002-6987-7834}\inst{\ref{aff109}}
\and A.~Secroun\orcid{0000-0003-0505-3710}\inst{\ref{aff66}}
\and G.~Seidel\orcid{0000-0003-2907-353X}\inst{\ref{aff22}}
\and E.~Sihvola\orcid{0000-0003-1804-7715}\inst{\ref{aff110}}
\and P.~Simon\inst{\ref{aff84}}
\and C.~Sirignano\orcid{0000-0002-0995-7146}\inst{\ref{aff5},\ref{aff103}}
\and G.~Sirri\orcid{0000-0003-2626-2853}\inst{\ref{aff32}}
\and L.~Stanco\orcid{0000-0002-9706-5104}\inst{\ref{aff103}}
\and P.~Tallada-Cresp\'{i}\orcid{0000-0002-1336-8328}\inst{\ref{aff41},\ref{aff42}}
\and A.~N.~Taylor\inst{\ref{aff48}}
\and H.~I.~Teplitz\orcid{0000-0002-7064-5424}\inst{\ref{aff111}}
\and I.~Tereno\orcid{0000-0002-4537-6218}\inst{\ref{aff59},\ref{aff112}}
\and N.~Tessore\orcid{0000-0002-9696-7931}\inst{\ref{aff56}}
\and S.~Toft\orcid{0000-0003-3631-7176}\inst{\ref{aff2},\ref{aff10}}
\and R.~Toledo-Moreo\orcid{0000-0002-2997-4859}\inst{\ref{aff113}}
\and F.~Torradeflot\orcid{0000-0003-1160-1517}\inst{\ref{aff42},\ref{aff41}}
\and I.~Tutusaus\orcid{0000-0002-3199-0399}\inst{\ref{aff114},\ref{aff115},\ref{aff106}}
\and L.~Valenziano\orcid{0000-0002-1170-0104}\inst{\ref{aff17},\ref{aff64}}
\and J.~Valiviita\orcid{0000-0001-6225-3693}\inst{\ref{aff78},\ref{aff79}}
\and T.~Vassallo\orcid{0000-0001-6512-6358}\inst{\ref{aff28},\ref{aff67}}
\and G.~Verdoes~Kleijn\orcid{0000-0001-5803-2580}\inst{\ref{aff1}}
\and A.~Veropalumbo\orcid{0000-0003-2387-1194}\inst{\ref{aff26},\ref{aff34},\ref{aff33}}
\and Y.~Wang\orcid{0000-0002-4749-2984}\inst{\ref{aff104}}
\and J.~Weller\orcid{0000-0002-8282-2010}\inst{\ref{aff69},\ref{aff68}}
\and F.~M.~Zerbi\orcid{0000-0002-9996-973X}\inst{\ref{aff26}}
\and E.~Zucca\orcid{0000-0002-5845-8132}\inst{\ref{aff17}}
\and M.~Huertas-Company\orcid{0000-0002-1416-8483}\inst{\ref{aff20},\ref{aff116},\ref{aff117}}
\and J.~Mart\'{i}n-Fleitas\orcid{0000-0002-8594-569X}\inst{\ref{aff118}}
\and V.~Scottez\orcid{0009-0008-3864-940X}\inst{\ref{aff119},\ref{aff120}}}
										   
\institute{Kapteyn Astronomical Institute, University of Groningen, PO Box 800, 9700 AV Groningen, The Netherlands\label{aff1}
\and
Cosmic Dawn Center (DAWN)\label{aff2}
\and
Space Telescope Science Institute, 3700 San Martin Dr, Baltimore, MD 21218, USA\label{aff3}
\and
INAF-Osservatorio Astronomico di Padova, Via dell'Osservatorio 5, 35122 Padova, Italy\label{aff4}
\and
Dipartimento di Fisica e Astronomia "G. Galilei", Universit\`a di Padova, Via Marzolo 8, 35131 Padova, Italy\label{aff5}
\and
Institute of Science and Technology Austria (ISTA), Am Campus 1, 3400 Klosterneuburg, Austria\label{aff6}
\and
Leiden Observatory, Leiden University, Einsteinweg 55, 2333 CC Leiden, The Netherlands\label{aff7}
\and
Department of Physical Sciences, Ritsumeikan University, Kusatsu, Shiga 525-8577, Japan\label{aff8}
\and
Academia Sinica Institute of Astronomy and Astrophysics (ASIAA), 11F of ASMAB, No.~1, Section 4, Roosevelt Road, Taipei 10617, Taiwan\label{aff9}
\and
Niels Bohr Institute, University of Copenhagen, Jagtvej 128, 2200 Copenhagen, Denmark\label{aff10}
\and
Centro de Astrobiolog\'ia (CAB), CSIC--INTA, Cra. de Ajalvir Km.~4, 28850 -- Torrej\'on de Ardoz, Madrid, Spain\label{aff11}
\and
Department of Mathematics and Physics, Roma Tre University, Via della Vasca Navale 84, 00146 Rome, Italy\label{aff12}
\and
INAF-Osservatorio Astronomico di Roma, Via Frascati 33, 00078 Monteporzio Catone, Italy\label{aff13}
\and
Kobayashi-Maskawa Institute for the Origin of Particles and the Universe, Nagoya University, Chikusa-ku, Nagoya, 464-8602, Japan\label{aff14}
\and
Institute for Particle Physics and Astrophysics, Dept. of Physics, ETH Zurich, Wolfgang-Pauli-Strasse 27, 8093 Zurich, Switzerland\label{aff15}
\and
Department of Astronomy, University of Geneva, ch. d'Ecogia 16, 1290 Versoix, Switzerland\label{aff16}
\and
INAF-Osservatorio di Astrofisica e Scienza dello Spazio di Bologna, Via Piero Gobetti 93/3, 40129 Bologna, Italy\label{aff17}
\and
Sterrenkundig Observatorium, Universiteit Gent, Krijgslaan 281 S9, 9000 Gent, Belgium\label{aff18}
\and
School of Physics \& Astronomy, University of Southampton, Highfield Campus, Southampton SO17 1BJ, UK\label{aff19}
\and
Instituto de Astrof\'{\i}sica de Canarias, E-38205 La Laguna, Tenerife, Spain\label{aff20}
\and
Universidad de La Laguna, Dpto. Astrof\'\i sica, E-38206 La Laguna, Tenerife, Spain\label{aff21}
\and
Max-Planck-Institut f\"ur Astronomie, K\"onigstuhl 17, 69117 Heidelberg, Germany\label{aff22}
\and
Department of Physics and Astronomy, University of British Columbia, Vancouver, BC V6T 1Z1, Canada\label{aff23}
\and
Jet Propulsion Laboratory, California Institute of Technology, 4800 Oak Grove Drive, Pasadena, CA, 91109, USA\label{aff24}
\and
ESAC/ESA, Camino Bajo del Castillo, s/n., Urb. Villafranca del Castillo, 28692 Villanueva de la Ca\~nada, Madrid, Spain\label{aff25}
\and
INAF-Osservatorio Astronomico di Brera, Via Brera 28, 20122 Milano, Italy\label{aff26}
\and
IFPU, Institute for Fundamental Physics of the Universe, via Beirut 2, 34151 Trieste, Italy\label{aff27}
\and
INAF-Osservatorio Astronomico di Trieste, Via G. B. Tiepolo 11, 34143 Trieste, Italy\label{aff28}
\and
INFN, Sezione di Trieste, Via Valerio 2, 34127 Trieste TS, Italy\label{aff29}
\and
SISSA, International School for Advanced Studies, Via Bonomea 265, 34136 Trieste TS, Italy\label{aff30}
\and
Dipartimento di Fisica e Astronomia, Universit\`a di Bologna, Via Gobetti 93/2, 40129 Bologna, Italy\label{aff31}
\and
INFN-Sezione di Bologna, Viale Berti Pichat 6/2, 40127 Bologna, Italy\label{aff32}
\and
Dipartimento di Fisica, Universit\`a di Genova, Via Dodecaneso 33, 16146, Genova, Italy\label{aff33}
\and
INFN-Sezione di Genova, Via Dodecaneso 33, 16146, Genova, Italy\label{aff34}
\and
Department of Physics "E. Pancini", University Federico II, Via Cinthia 6, 80126, Napoli, Italy\label{aff35}
\and
INAF-Osservatorio Astronomico di Capodimonte, Via Moiariello 16, 80131 Napoli, Italy\label{aff36}
\and
Dipartimento di Fisica, Universit\`a degli Studi di Torino, Via P. Giuria 1, 10125 Torino, Italy\label{aff37}
\and
INFN-Sezione di Torino, Via P. Giuria 1, 10125 Torino, Italy\label{aff38}
\and
INAF-Osservatorio Astrofisico di Torino, Via Osservatorio 20, 10025 Pino Torinese (TO), Italy\label{aff39}
\and
INAF-IASF Milano, Via Alfonso Corti 12, 20133 Milano, Italy\label{aff40}
\and
Centro de Investigaciones Energ\'eticas, Medioambientales y Tecnol\'ogicas (CIEMAT), Avenida Complutense 40, 28040 Madrid, Spain\label{aff41}
\and
Port d'Informaci\'{o} Cient\'{i}fica, Campus UAB, C. Albareda s/n, 08193 Bellaterra (Barcelona), Spain\label{aff42}
\and
Institute for Theoretical Particle Physics and Cosmology (TTK), RWTH Aachen University, 52056 Aachen, Germany\label{aff43}
\and
Deutsches Zentrum f\"ur Luft- und Raumfahrt e. V. (DLR), Linder H\"ohe, 51147 K\"oln, Germany\label{aff44}
\and
INFN section of Naples, Via Cinthia 6, 80126, Napoli, Italy\label{aff45}
\and
Institute for Astronomy, University of Hawaii, 2680 Woodlawn Drive, Honolulu, HI 96822, USA\label{aff46}
\and
Dipartimento di Fisica e Astronomia "Augusto Righi" - Alma Mater Studiorum Universit\`a di Bologna, Viale Berti Pichat 6/2, 40127 Bologna, Italy\label{aff47}
\and
Institute for Astronomy, University of Edinburgh, Royal Observatory, Blackford Hill, Edinburgh EH9 3HJ, UK\label{aff48}
\and
Jodrell Bank Centre for Astrophysics, Department of Physics and Astronomy, University of Manchester, Oxford Road, Manchester M13 9PL, UK\label{aff49}
\and
European Space Agency/ESRIN, Largo Galileo Galilei 1, 00044 Frascati, Roma, Italy\label{aff50}
\and
Universit\'e Claude Bernard Lyon 1, CNRS/IN2P3, IP2I Lyon, UMR 5822, Villeurbanne, F-69100, France\label{aff51}
\and
Institut de Ci\`{e}ncies del Cosmos (ICCUB), Universitat de Barcelona (IEEC-UB), Mart\'{i} i Franqu\`{e}s 1, 08028 Barcelona, Spain\label{aff52}
\and
Instituci\'o Catalana de Recerca i Estudis Avan\c{c}ats (ICREA), Passeig de Llu\'{\i}s Companys 23, 08010 Barcelona, Spain\label{aff53}
\and
Institut de Ciencies de l'Espai (IEEC-CSIC), Campus UAB, Carrer de Can Magrans, s/n Cerdanyola del Vall\'es, 08193 Barcelona, Spain\label{aff54}
\and
UCB Lyon 1, CNRS/IN2P3, IUF, IP2I Lyon, 4 rue Enrico Fermi, 69622 Villeurbanne, France\label{aff55}
\and
Mullard Space Science Laboratory, University College London, Holmbury St Mary, Dorking, Surrey RH5 6NT, UK\label{aff56}
\and
Canada-France-Hawaii Telescope, 65-1238 Mamalahoa Hwy, Kamuela, HI 96743, USA\label{aff57}
\and
Aix-Marseille Universit\'e, CNRS, CNES, LAM, Marseille, France\label{aff58}
\and
Departamento de F\'isica, Faculdade de Ci\^encias, Universidade de Lisboa, Edif\'icio C8, Campo Grande, PT1749-016 Lisboa, Portugal\label{aff59}
\and
Instituto de Astrof\'isica e Ci\^encias do Espa\c{c}o, Faculdade de Ci\^encias, Universidade de Lisboa, Campo Grande, 1749-016 Lisboa, Portugal\label{aff60}
\and
Universit\'e Paris-Saclay, CNRS, Institut d'astrophysique spatiale, 91405, Orsay, France\label{aff61}
\and
INAF-Istituto di Astrofisica e Planetologia Spaziali, via del Fosso del Cavaliere, 100, 00100 Roma, Italy\label{aff62}
\and
Space Science Data Center, Italian Space Agency, via del Politecnico snc, 00133 Roma, Italy\label{aff63}
\and
INFN-Bologna, Via Irnerio 46, 40126 Bologna, Italy\label{aff64}
\and
School of Physics, HH Wills Physics Laboratory, University of Bristol, Tyndall Avenue, Bristol, BS8 1TL, UK\label{aff65}
\and
Aix-Marseille Universit\'e, CNRS/IN2P3, CPPM, Marseille, France\label{aff66}
\and
University Observatory, LMU Faculty of Physics, Scheinerstr.~1, 81679 Munich, Germany\label{aff67}
\and
Max Planck Institute for Extraterrestrial Physics, Giessenbachstr. 1, 85748 Garching, Germany\label{aff68}
\and
Universit\"ats-Sternwarte M\"unchen, Fakult\"at f\"ur Physik, Ludwig-Maximilians-Universit\"at M\"unchen, Scheinerstr.~1, 81679 M\"unchen, Germany\label{aff69}
\and
Institute of Theoretical Astrophysics, University of Oslo, P.O. Box 1029 Blindern, 0315 Oslo, Norway\label{aff70}
\and
Department of Physics, Lancaster University, Lancaster, LA1 4YB, UK\label{aff71}
\and
Felix Hormuth Engineering, Goethestr. 17, 69181 Leimen, Germany\label{aff72}
\and
Technical University of Denmark, Elektrovej 327, 2800 Kgs. Lyngby, Denmark\label{aff73}
\and
Cosmic Dawn Center (DAWN), Denmark\label{aff74}
\and
NASA Goddard Space Flight Center, Greenbelt, MD 20771, USA\label{aff75}
\and
Department of Physics and Astronomy, University College London, Gower Street, London WC1E 6BT, UK\label{aff76}
\and
Universit\'e de Gen\`eve, D\'epartement de Physique Th\'eorique and Centre for Astroparticle Physics, 24 quai Ernest-Ansermet, CH-1211 Gen\`eve 4, Switzerland\label{aff77}
\and
Department of Physics, P.O. Box 64, University of Helsinki, 00014 Helsinki, Finland\label{aff78}
\and
Helsinki Institute of Physics, Gustaf H{\"a}llstr{\"o}min katu 2, University of Helsinki, 00014 Helsinki, Finland\label{aff79}
\and
Laboratoire d'etude de l'Univers et des phenomenes eXtremes, Observatoire de Paris, Universit\'e PSL, Sorbonne Universit\'e, CNRS, 92190 Meudon, France\label{aff80}
\and
SKAO, Jodrell Bank, Lower Withington, Macclesfield SK11 9FT, UK\label{aff81}
\and
Centre de Calcul de l'IN2P3/CNRS, 21 avenue Pierre de Coubertin 69627 Villeurbanne Cedex, France\label{aff82}
\and
University of Applied Sciences and Arts of Northwestern Switzerland, School of Computer Science, 5210 Windisch, Switzerland\label{aff83}
\and
Universit\"at Bonn, Argelander-Institut f\"ur Astronomie, Auf dem H\"ugel 71, 53121 Bonn, Germany\label{aff84}
\and
INFN-Sezione di Roma, Piazzale Aldo Moro, 2 - c/o Dipartimento di Fisica, Edificio G. Marconi, 00185 Roma, Italy\label{aff85}
\and
Dipartimento di Fisica e Astronomia "Augusto Righi" - Alma Mater Studiorum Universit\`a di Bologna, via Piero Gobetti 93/2, 40129 Bologna, Italy\label{aff86}
\and
Department of Physics, Institute for Computational Cosmology, Durham University, South Road, Durham, DH1 3LE, UK\label{aff87}
\and
Universit\'e Paris Cit\'e, CNRS, Astroparticule et Cosmologie, 75013 Paris, France\label{aff88}
\and
CNRS-UCB International Research Laboratory, Centre Pierre Bin\'etruy, IRL2007, CPB-IN2P3, Berkeley, USA\label{aff89}
\and
Institute of Physics, Laboratory of Astrophysics, Ecole Polytechnique F\'ed\'erale de Lausanne (EPFL), Observatoire de Sauverny, 1290 Versoix, Switzerland\label{aff90}
\and
Telespazio UK S.L. for European Space Agency (ESA), Camino bajo del Castillo, s/n, Urbanizacion Villafranca del Castillo, Villanueva de la Ca\~nada, 28692 Madrid, Spain\label{aff91}
\and
Institut de F\'{i}sica d'Altes Energies (IFAE), The Barcelona Institute of Science and Technology, Campus UAB, 08193 Bellaterra (Barcelona), Spain\label{aff92}
\and
School of Mathematics and Physics, University of Surrey, Guildford, Surrey, GU2 7XH, UK\label{aff93}
\and
European Space Agency/ESTEC, Keplerlaan 1, 2201 AZ Noordwijk, The Netherlands\label{aff94}
\and
School of Mathematics, Statistics and Physics, Newcastle University, Herschel Building, Newcastle-upon-Tyne, NE1 7RU, UK\label{aff95}
\and
DARK, Niels Bohr Institute, University of Copenhagen, Jagtvej 155, 2200 Copenhagen, Denmark\label{aff96}
\and
Waterloo Centre for Astrophysics, University of Waterloo, Waterloo, Ontario N2L 3G1, Canada\label{aff97}
\and
Department of Physics and Astronomy, University of Waterloo, Waterloo, Ontario N2L 3G1, Canada\label{aff98}
\and
Perimeter Institute for Theoretical Physics, Waterloo, Ontario N2L 2Y5, Canada\label{aff99}
\and
Universit\'e Paris-Saclay, Universit\'e Paris Cit\'e, CEA, CNRS, AIM, 91191, Gif-sur-Yvette, France\label{aff100}
\and
Centre National d'Etudes Spatiales -- Centre spatial de Toulouse, 18 avenue Edouard Belin, 31401 Toulouse Cedex 9, France\label{aff101}
\and
Institute of Space Science, Str. Atomistilor, nr. 409 M\u{a}gurele, Ilfov, 077125, Romania\label{aff102}
\and
INFN-Padova, Via Marzolo 8, 35131 Padova, Italy\label{aff103}
\and
Caltech/IPAC, 1200 E. California Blvd., Pasadena, CA 91125, USA\label{aff104}
\and
Instituto de F\'isica Te\'orica UAM-CSIC, Campus de Cantoblanco, 28049 Madrid, Spain\label{aff105}
\and
Institut de Recherche en Astrophysique et Plan\'etologie (IRAP), Universit\'e de Toulouse, CNRS, UPS, CNES, 14 Av. Edouard Belin, 31400 Toulouse, France\label{aff106}
\and
Universit\'e St Joseph; Faculty of Sciences, Beirut, Lebanon\label{aff107}
\and
Departamento de F\'isica, FCFM, Universidad de Chile, Blanco Encalada 2008, Santiago, Chile\label{aff108}
\and
Universit\"at Innsbruck, Institut f\"ur Astro- und Teilchenphysik, Technikerstr. 25/8, 6020 Innsbruck, Austria\label{aff109}
\and
Department of Physics and Helsinki Institute of Physics, Gustaf H\"allstr\"omin katu 2, University of Helsinki, 00014 Helsinki, Finland\label{aff110}
\and
Infrared Processing and Analysis Center, California Institute of Technology, Pasadena, CA 91125, USA\label{aff111}
\and
Instituto de Astrof\'isica e Ci\^encias do Espa\c{c}o, Faculdade de Ci\^encias, Universidade de Lisboa, Tapada da Ajuda, 1349-018 Lisboa, Portugal\label{aff112}
\and
Universidad Polit\'ecnica de Cartagena, Departamento de Electr\'onica y Tecnolog\'ia de Computadoras,  Plaza del Hospital 1, 30202 Cartagena, Spain\label{aff113}
\and
Institute of Space Sciences (ICE, CSIC), Campus UAB, Carrer de Can Magrans, s/n, 08193 Barcelona, Spain\label{aff114}
\and
Institut d'Estudis Espacials de Catalunya (IEEC),  Edifici RDIT, Campus UPC, 08860 Castelldefels, Barcelona, Spain\label{aff115}
\and
Universit\'e PSL, Observatoire de Paris, Sorbonne Universit\'e, CNRS, LERMA, 75014, Paris, France\label{aff116}
\and
Universit\'e Paris-Cit\'e, 5 Rue Thomas Mann, 75013, Paris, France\label{aff117}
\and
Aurora Technology for European Space Agency (ESA), Camino bajo del Castillo, s/n, Urbanizacion Villafranca del Castillo, Villanueva de la Ca\~nada, 28692 Madrid, Spain\label{aff118}
\and
Institut d'Astrophysique de Paris, 98bis Boulevard Arago, 75014, Paris, France\label{aff119}
\and
ICL, Junia, Universit\'e Catholique de Lille, LITL, 59000 Lille, France\label{aff120}}    


 
  \abstract{Little Red Dots (LRDs) are some the most intriguing galaxy populations recently identified at $z \gsim 4$ with JWST. They constitute the most extreme class of a more abundant population of sources with `V-shaped' spectral energy distributions (SEDs) and compact morphologies, which includes also Little Blue Dots (LBDs).  Finding brighter analogues to these sources requires surveying sky areas which are significantly larger than those covered with JWST. \Euclid deep images are ideally suited for this purpose. We make use of \Euclid near-infrared images, complemented by \textit{Spitzer} Infrared Array Camera (IRAC) data, over 0.75~$\rm deg^2$ of the COSMOS field to select a sample of 233 sources with `V-shaped' SEDs at $z>4$. Out of those, we identify 16 sources with compactness $>1\sigma$ above the median of all $z>4$ galaxies, which we consider robust LRD/LBD candidates in our sample. The stellar masses of these 16 sources are in the range  $\rm [10^{8.5},10^{10.5}] \it \, M_\odot$, so they are significantly more massive than typical JWST-selected LRDs/LBDs. Interestingly, half of them are about as old as the Universe at their redshifts.   In addition, we find that the median photometric properties of the \Euclid LRDs/LBDs are similar to those of the so-called Blue Dust-Obscured Galaxies (Blue DOGs). Less than 10\% of all our `V-shaped' SED sources, including only one of the \Euclid LBDs, correspond to known AGN. The latter mostly constitute a population disjoint to the `V-shaped' SED sources. Spectroscopic follow up of the \Euclid LRDs/LBD candidates remains necessary to probe whether they host BLAGN as fainter analogues do and whether constitute a transition phase from these fainter sources to standard AGN.
  }

   
    \let\footnote\thanks
   \maketitle

%

\section{Introduction} 


Understanding how the oldest  galaxies that we see in the Universe today formed and evolved is one of the most important questions of extragalactic astronomy.  In particular, it is most relevant to investigate their properties at high redshifts, when galaxies were still young and passing by their early evolutionary stages. And, while substantial knowledge has been achieved thanks to the large number of studies conducted over the past few decades, exploring the connection between early galaxy populations is still largely work in progress.

The study of the high-redshift Universe is now in a golden decade thanks to the advent of major infrared (IR) space telescopes. One of them is \Euclid \citep{EuclidSkyOverview}, which is obtaining relatively deep near-IR imaging and spectroscopy over unprecedentedly large areas of the sky. \Euclid will carry out two main galaxy surveys, one Wide and one Deep, with the latter reaching a depth of 26.4~mag (5$\sigma$)  over more than 50~deg$^{2}$ after five years of operation \citep{Scaramella-EP1}. The Euclid Deep Survey \citep{EP-McPartland,Zalesky_2025} will allow us to probe galaxies down to stellar mass $M_* \approx 10^{8} \, \it{M_\odot}$ up to cosmic noon ($z \approx 3$), and the rarest and brightest galaxies well into the epoch of reionisation \citep[$z>6$; e.g., ][]{NavarroCarrera25_Euclid}.

The other current main IR space telescope complementing \Euclid is JWST. The latter is providing much deeper data (by up to about 5~mag), albeit in much smaller areas of the sky. At high redshifts ($z>4$), JWST observations are enabling the discovery of faint galaxy populations that are beyond the observing possibilities of almost any other currently operating observatory. 

 One of the most intriguing classes of objects that has been discovered by JWST at high $z$  are the so-called `Little Red Dots' \citep[LRDs; ][]{Labbe_2023a, Matthee_2024, Kokorev_2024}. These are very compact sources at rest-frame optical wavelengths, which are characterised by a `V-shape' in their $f_\lambda (\lambda)$ spectral energy distribution (SED), displaying red colours at rest-frame wavelengths $\lambda_{\rm rest} \in [3645,5000] \, \rm \text{\AA}$, 
 and flat or blue colours at rest-frame $\lambda_{\rm rest} \in [1000,2000] \, \rm \text{\AA}$ \citep{Furtak_2023,Kocevski_2024}. By performing the identification based on JWST Near Infrared Camera (NIRCam) band colours,  the vast majority of the selected sources lie naturally at $z>4$ \citep{Kokorev_2024,Kocevski_2024,Ma_2025}. Spectroscopic follow up has shown that the vast majority of LRDs host a broad-line active galactic nucleus \citep[BLAGN; e.g,][]{Hviding2025}

 However, LRDs are only a minority of all the faint and compact BLAGN discovered with JWST at high $z$ \citep{Hainline_2025}. The others are bluer and, thus, by extension have been named Little Blue Dots \citep[LBDs;][]{Brazzini2026}. In a recent work \citet{MM26} argue that LRDs and LBDs constitute in essence a common and continuous class of objects, which mainly differ by being observed at different orientation angles. Even within LRD populations, studies have found that they are a spectroscopically diverse population, spanning a continuous range of properties \citep{Perez_Gonzalez_2026,Rinaldi_2026}.  

In any case, LRDs and LBDs appear to be different to standard AGN for a number of reasons. In particular, the inferred central black holes (BH) masses, as derived from the Balmer-line broadening in their spectra,  appear to be relatively high with respect to those of their hosts, i.e. these sources lie significantly above the local $M_{\rm BH}$--$M_{\rm \ast}$ relation \citep{Harikane2023,Maiolino_2025}, although this argument is still under debate \citep[e.g.,][]{Rusakov26}. It is unclear how such presumably overmassive BH could have formed at early cosmic times, but it is possible that their growth has been facilitated by dense star clusters \citep{Caputi2026, KS2026}. 

 In general, no extended stellar component is detected around LRDs, suggesting that the corresponding host galaxies must have had relatively low stellar masses. However, more recent studies have revealed that, in spite of the compactness, LRD morphology can actually be complex \citep{Rinaldi_2024}, showing signatures of galaxy mergers in some cases. In addition, some high-$z$ sources have been found to present SEDs with similar photometric properties to LRDs, in spite of not being compact \citep{Rinaldi_2025Virgil, Iani_2025}.
 
Furthermore, recent observations at $z<4$ have revealed galaxies with extended structures hosting compact, LRD-like nuclei \citep{Billand_2025,Rinaldi_2025b}. These results suggest that the characteristic `V-shaped' double power-law SED is not exclusive to LRDs. Other newer studies at low-$z$ have yielded discoveries of rare LRDs in the local Universe ($z=0.1-0.2$), with properties consistent with their high-$z$ counterparts \citep{Lin_2025, Iani2026}.  Some of these local LRDs are also found to have compact yet clumpy host galaxies.


In fact, very luminous examples of compact objects with V-shaped SEDs at $z\approx $ 2--3 have been known since many years before the JWST launch. These are the so-called Blue-Excess Dust-Obscured Galaxies \citep[BlueDOGs;][]{Assef_2016,Noboriguchi_2019}, which constitute about 1\% of all DOGs \citep{Dey_2008, Pope_2008}. BlueDOGs are thought to be dust-obscured quasars which have just expelled their dust and are in transition to the optically thin quasar phase. While for many years BlueDOGs have been considered little more than a curiosity, JWST's discovery of LRDs has reignited the interest in these sources \citep[e.g.,][]{Kim_2025}. The exact evolutionary connection between BlueDOGs and LRDs, if any, is unclear, though. BlueDOGs are several dex more luminous than typical JWST high-$\it z$ sources. The \Euclid sources that we select and analyse here could potentially constitute the bridge between those galaxy populations.

The Euclid Deep Survey near-IR imaging, complemented with  \Spitzer Infrared Array Camera \citep{Fazio_2004} imaging at 3.6 and 4.5$\, \rm \micron$, will be ideal to investigate the existence of a population of sources with V-shaped SEDs at high redshifts, which are scaled-up with respect to the JWST-discovered LRDs. However, the Euclid Deep Survey will only reach its expected final depth ($\approx 26.5 \,$mag at wavelengths 1.0--1.6$\, \micron$) in some years from now. At the moment of writing, the depth of the \Euclid data in the Deep Survey fields only allows for the study of very rare high-$z$ objects.

Fortunately, \Euclid has already collected data of such depth in the so-called  auxiliary fields. Although these regions of the sky are much smaller than the fields comprising the  Euclid Deep Survey,  the \Euclid imaging  already has comparable depth to that of the Euclid Deep Survey final depth. 

Here we present an investigation of V-shaped double power-law SED sources selected using \Euclid data in COSMOS  \citep{Scoville_2007}, which is  one of \textit{Euclid'}s auxiliary fields.  A priori we do not put any restriction on their compactness, but rather study their properties for different compactness values. In a second stage, we do segregate the most compact sources and consider them the most robust candidates to scaled-up LRD/LBDs in our sample. Our goal is to study them in the same context as the JWST-selected faint AGN and standard AGN.  We focus exclusively on galaxies at $z>4$, while the analysis of \textit{Euclid}-selected V-shaped SED sources at lower redshifts has already been presented elsewhere \citep{Q1-SP011}. 

The layout of this paper is as follows.  In Sect. \ref{sec:data} we describe the \textit{Euclid} data products and ancillary data in COSMOS. Then, in Sect.~\ref{sec:photometry} we describe how we performed our photometric measurements and SED-fitting. In Sect.~\ref{sec:sample} we lay out our source selection methodology. In Sect.~\ref{sec:dplprop} we analyse the basic properties of the V-shaped SED selected sources. In Sect.~\ref{sec:agn},  we show that classical AGN are mostly not included in the V-shaped SED selection. In Sect. \ref{sec:lrd_candidates}, we identify scaled-up LRDs/LBDs within our sample, based on their compactness, and place them in the same physical framework as the JWST-selected sources by calculating their rest-frame UV luminosity function. Our aim is to assess whether they could constitute a luminous extension of the JWST LRD/LBD population. We explain the differences between the JWST and \Euclid V-shaped SED source selection in Sect.~\ref{sec:disc}. In Sect.~\ref{sec:bluedogs}, we discuss the compared properties of our sources and those of BlueDOGs. Finally, in Sect.~\ref{sec:conclusion} we summarise our findings and present general conclusions. We adopt a $\Lambda$CDM cosmology with $H_0=70 \, \rm km \, s^{-1} \, Mpc^{-1}$, $\Omega_{\rm m} = 0.3$, $\Omega_{\rm \Lambda}$ = 0.7. All magnitudes refer to the AB system \citep{oke_83}. We consider throughout a Chabrier initial mass function \citep{Chabrier_2003} for all stellar masses and star-formation histories.

\section{Datasets} \label{sec:data}

In this work we analyse the \Euclid images covering about 0.75~deg$^2$ of the COSMOS field, in four photometric bands, namely $\IE$ from the Visible Camera (VIS) instrument \citep{EuclidSkyVIS}, and the NIR $\YE, \JE$, and $\HE$, from the Near Infrared Spectrometer and Photometer (NISP) instrument \citep{EuclidSkyNISP}. 

To allow for a wider photometric coverage, we complement these datasets by also using the COSMOS2020 photometric catalogue \citep{Weaver_2022}, which covers an area of 2 deg$^2$ and contains photometric data in the UV and optical from ground-based observations with multiple instruments, spanning $[0.3,0.9] \, \micron$, as well as near- and mid-infrared observations with the VISTA $K_{\rm s}$ band \citep{McCracken_2012} and \Spitzer IRAC \citep{Moneti-EP17}. Here we consider specifically 25 filters from COSMOS2020: CFHT MegaCam $u^*$, Subaru Suprime-Cam \emph{B}, \emph{V}, \emph{$r^+$}, \emph{$i^+$}, \emph{$z^+$}, \emph{$z^{++}$}; 14 intermediate- and narrow-bands; VISTA $K_{\rm s}$; and \textit{Spitzer} IRAC channels 1 and 2 (IRAC1 and IRAC2). The full list of filters used, including the \Euclid bands, is quoted in Table~\ref{table:litstuff}. From COSMOS2020 we use the \texttt{Classic} photometric catalogue version, which was built after measuring IRAC photometry with the PSF-fitting algorithm IRACLEAN \citep{Hsieh_2012} --  see \citep{Weaver_2022} for details.

\begin{table}
\centering
\caption{List of filters used for SED fitting in this work. Columns: instrument; band; central wavelength, $\lambda_{\rm c}$; bandwidth, FWHM;  and 3$\sigma$ depth in AB magnitudes.}
\label{table:litstuff}
\footnotesize
\resizebox{\columnwidth}{!}{%
\begin{tabular}{p{3.1cm} c r r r}
\toprule
Instrument & Band & $\lambda_{\rm c}$ & Width & Depth \\
  &   & [\si{\angstrom}] & [\si{\angstrom}] & [mag] \\
\midrule
MegaCam/CFHT & $u^*$ & 3858 & 598 & 27.7 \\
ACS/HST & F814W & 8333 & 2511 & 27.8 \\
Suprime-Cam/Subaru\tablefootmark{a} & $B$ & 4488 & 892 & 27.8 \\
" & $V$ & 5487 & 954 & 26.8 \\
" & $r^+$ & 6305 & 1376 & 27.1 \\
" & $i^+$ & 7693 & 1497 & 26.7 \\
" & $z^+$ & 8973 & 847 & 25.7 \\
" & $z^{++}$ & 9063 & 1335 & 26.3 \\
" & $IB427$ & 4266 & 207 & 26.1 \\
" & $IB464$ & 4635 & 218 & 25.6 \\
" & $IA484$ & 4851 & 229 & 26.5 \\
" & $IB505$ & 5064 & 231 & 26.1 \\
" & $IA527$ & 5064 & 231 & 26.1 \\
" & $IB505$ & 5261 & 243 & 26.4 \\
" & $IB574$ & 5766 & 273 & 25.8 \\
" & $IA624$ & 6232 & 300 & 26.4 \\
" & $IA679$ & 6780 & 336 & 25.6 \\
" & $IB709$ & 7073 & 316 & 25.9 \\
" & $IA738$ & 7361 & 324 & 26.1 \\
" & $IA767$ & 7694 & 365 & 25.6 \\
" & $IB827$ & 8243 & 343 & 25.6 \\
" & $NB711$ & 7121 & 72 & 25.5 \\
" & $NB816$ & 8150 & 120 & 25.6 \\
VIS/\Euclid & \IE & 7103 & 3318 & 28.0 \\
NISP/\Euclid & \YE & 10\,809 & 2627 & 26.8 \\
" & \JE & 13\,673 & 3994 & 26.8 \\
" & \HE & 17\,714 & 4999 & 26.7 \\
VIRCAM/VISTA\tablefootmark{b} & $K_{\rm s}$ & 21\,557 & 3074 & 25.7 \\
IRAC/\textit{Spitzer}\tablefootmark{c,d,e} & ch1 & 35\,686 & 7443 & 26.4 \\
" & ch2 & 45\,067 & 10\,119 & 26.3 \\
\bottomrule
\end{tabular}}
\tablefoot{
\tablefoottext{a}{\citet{Taniguchi_2007}};
\tablefoottext{b}{\citet{McCracken_2012}};
\tablefoottext{c}{\citet{Sanders_2003}};
\tablefoottext{d}{\citet{Capak_2013}};
\tablefoottext{e}{\citet{Ashby_2018}}
}
\end{table}

\section{Photometric measurements and analysis Techniques}\label{sec:photometry}

\subsection{Source detection and photometry}

To measure photometry on the \Euclid images, we used the \texttt{SourceExtractor} software \citep{Bertin_Arnouts_1996}\, running it in dual mode with the $\YE$, $\JE$, and $\HE$ images stacked together as the detection image. We considered the root mean squared (RMS) images for each filter as inverse weight maps, and set the detection threshold to 1.5$\sigma$ over 3 contiguous pixels. We choose such low detection threshold to maximize the number of detected sources. We validate the reliability of those sources by cross-matching the \Euclid catalogue with the COSMOS2020 one. 

For the \Euclid $\IE$, $\YE$, $\JE$, and $\HE$ images, we used the zero-point magnitudes 24.4, 29.8, 30.0, and 30.0 respectively. We measured photometry in circular apertures of $\ang{; ;2}$ diameter. To obtain the flux corrections, we took the ratio of the fluxes within a $\ang{; ;10}$ aperture and a $\ang{; ;2}$ aperture for ten non-saturated stars, then averaged the values.  Following that, to account for Galactic extinction, we utilised the \texttt{dustmaps} python module with the dust maps from \cite{Schlafly_Finkbeiner_2011}. 

The initial \Euclid catalogue contained 648\,752 sources. After cross-matching it with the COSMOS2020 catalogue adopting a matching radius of $\ang{; ;0.5}$ we obtained a combined catalogue with 357\,011 sources. To clean for bright stars we removed the point-like sources (\texttt{SExtractor} CLASS\_STAR parameter $\geq$ 0.9) that displayed the characteristic colours of Galactic stars, which appear segregated in the ($B-\JE$) and ($\JE-[3.6]$) colour-colour diagram \citep{Caputi_2011,Caputi_2015}. We also removed any objects whose \Euclid $\YE$, $\JE$, and $\HE$ magnitudes were much brighter than the corresponding UVISTA $Y$, $J$, and $H$ magnitudes from in the catalogue by \cite{Weaver_2022}, indicating that their photometry is likely contaminated. Finally, we also removed sources with $<$17 mag in any band, which also likely have contaminated photometry. 
Removing these suspicious sources resulted in a final catalogue of 287\,919 sources.

\subsection{Galaxy physical parameters from SED fitting}

\begin{figure}
    \centering
    \includegraphics[width=0.45\textwidth]{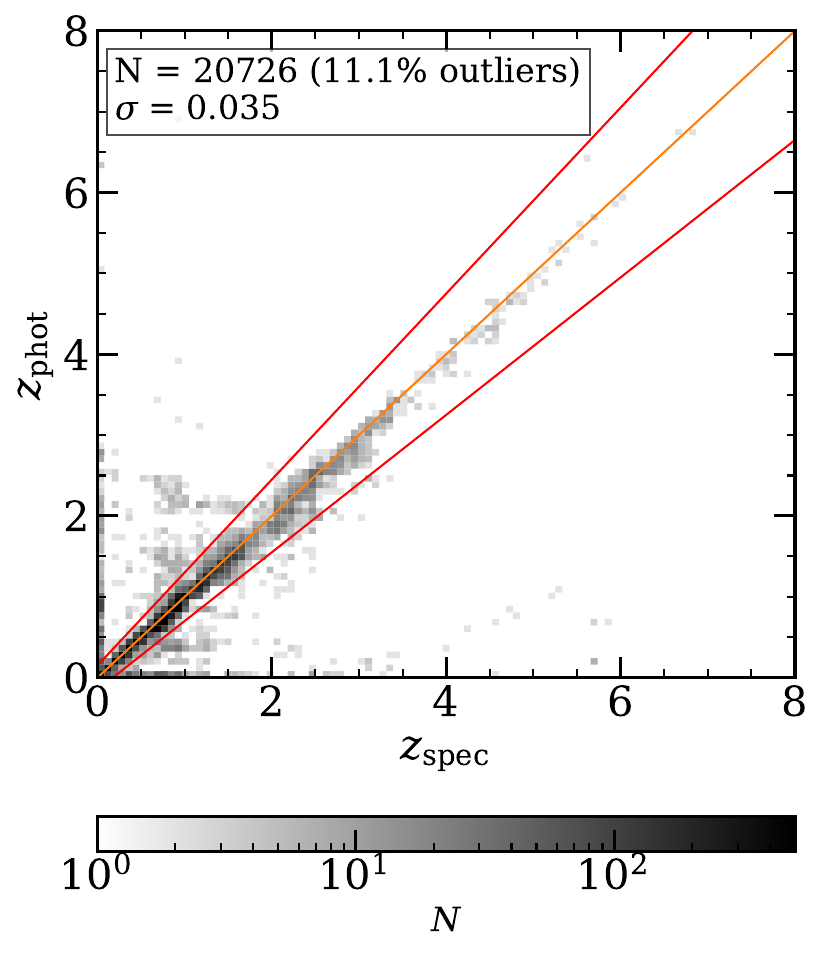}
\caption{Photometric redshifts obtained from SED fitting with \texttt{LePhare} versus the spectroscopic redshifts from \cite{Khostovan_2025} catalogue.  The identity line is shown in orange, and the red lines denote the upper and lower bounds, beyond which are the catastrophic outliers, defined as $|z_{\rm spec}$$-z_{\rm phot}|/(1+z_{\rm spec}) > 0.15$. From the 20\,726 galaxies in this diagnostic, we have 11.1\% outliers, with a dispersion of $\sigma=0.035$.}
    \label{fig:zphot_zspec}
\end{figure}

Using the combined \Euclid-COSMOS2020 photometric catalogue,  we performed SED fitting to obtain photometric redshifts, $z_{\rm phot}$, and derive main physical parameters using the $\chi^2$ fitting algorithm \texttt{LePHARE} \citep{Arnouts_Ilbert_2011}. We adopted a similar methodology to \cite{vanMierlo-EP21} and used 18 galaxy models \cite[][BC03]{Bruzual_Charlot_2003}, with exponentially declining star-formation histories (SFHs) at timescales $\tau$ = 0.01, 0.1, 0.3, 1.0, 3.0, 5.0, 10.0, and 15~Gyr. We made use of two metallicities, one solar ($Z = \rm Z_\odot$) and one sub-solar ($Z = 0.2 \, Z_\odot$). We ran the SED fitting over the redshift range $z  \in [0,8]$ with step $\rm d \it z \rm = 0.04$.

To  take into account the effect of dust attenuation, we convolved the galaxy templates using the \cite{Calzetti_2000} and \cite{Leitherer_2002} reddening law, allowing for the colour excess to vary as a free parameter within the range $E(B-V)\in [0,1]$, with increments of 0.1. We also allowed \texttt{LePHARE} to add emission lines to the templates. 

As it has been common practice in most works based on the COSMOS catalogues \citep{Dahlen_2013},  we multiplied the flux errors by a factor of 1.5. Because the IRAC1 and IRAC2 flux errors were underestimated, we imposed a minimum error corresponding to a signal-to-noise ratio $\rm S/N$ = 10 for these bands to improve the quality of the SED fitting. Without imposing minimum errors, the photometric redshifts for some of the sources would be overestimated, causing spurious $z_{\rm phot}$ columns of sources at $z\sim4, 5, 6,$ and 8 on the $z_{\rm phot}$--$z_{\rm spec}$ plane. To account for non-detections, we substituted them with the $3 \sigma$ flux upper limits in each of their bands, selecting the option in \texttt{LePHARE} which rejects SED templates with fluxes higher than the $3 \sigma$ upper limits in the non-detection bands. We then computed photometric zero-point corrections to improve the quality of the fittings, by calculating the median of the offsets between the observed and model fluxes in each band, and applying them onto the photometric catalogue \citep{Deshmukh_2018,vanMierlo-EP21}. This process was repeated until the offsets  and the improvement in the outlier diagnostic converged.

To check the quality of the fittings, we compared the output $z_{\rm phot}$ with the spectroscopic redshifts, $z_{\rm spec}$, available for COSMOS. We considered the $z_{\rm spec}$ compilation from \cite{Khostovan_2025}. Out of our 287\,919 sources, 20\,726 had a counterpart in the \cite{Khostovan_2025} catalogue. To assess the quality of the photometric redshifts, we analysed the distribution of the metric $|z_{\rm spec}-z_{\rm phot}|/(1+z_{\rm spec})$. The results can be seen in Fig.\ref{fig:zphot_zspec}. The diagnostic distribution has a small dispersion ($\sigma=0.035$) with 11.1\% catastrophic outliers.

In addition, to test the validity of the $z > 4$ solutions, we checked whether any of the high-redshift candidate galaxies has detections in filters bluewards of the Lyman break. Following slightly modified criteria from \cite{vanMierlo-EP21} and \cite{Caputi_2015}, we discarded sources with $z_{\rm phot} \in [4,4.5]$ and a $>2\sigma$ detection in the $U$ band; $z_{\rm phot} \in [4.5,6]$ whilst having a $>2 \sigma$ detection in the $B$ band;  sources with $z_{\rm phot} \in [6,6.7]$  along with a $>2\sigma$ detection in the $V$ band; and also sources with $z > 7$ with a $>2 \sigma$ detection in the $i$ band. This cleaning of the output redshift catalogue at $z>4$ removed $\sim$\,12\% of the high-redshift candidates. 

For the sources that had spectroscopic redshift data available, we replaced their photometric redshifts with the $z_{\rm spec}$ values from the catalogue by \cite{Khostovan_2025}. Ultimately, our final high-redshift catalogue contains $6451$ sources at $z>4$, which are 2.2\% of the \Euclid-COSMOS catalogue.

\subsection{AGN identification}\label{sec:agn_identification}

\begin{figure*}
    \centering
    \includegraphics[width=\textwidth]{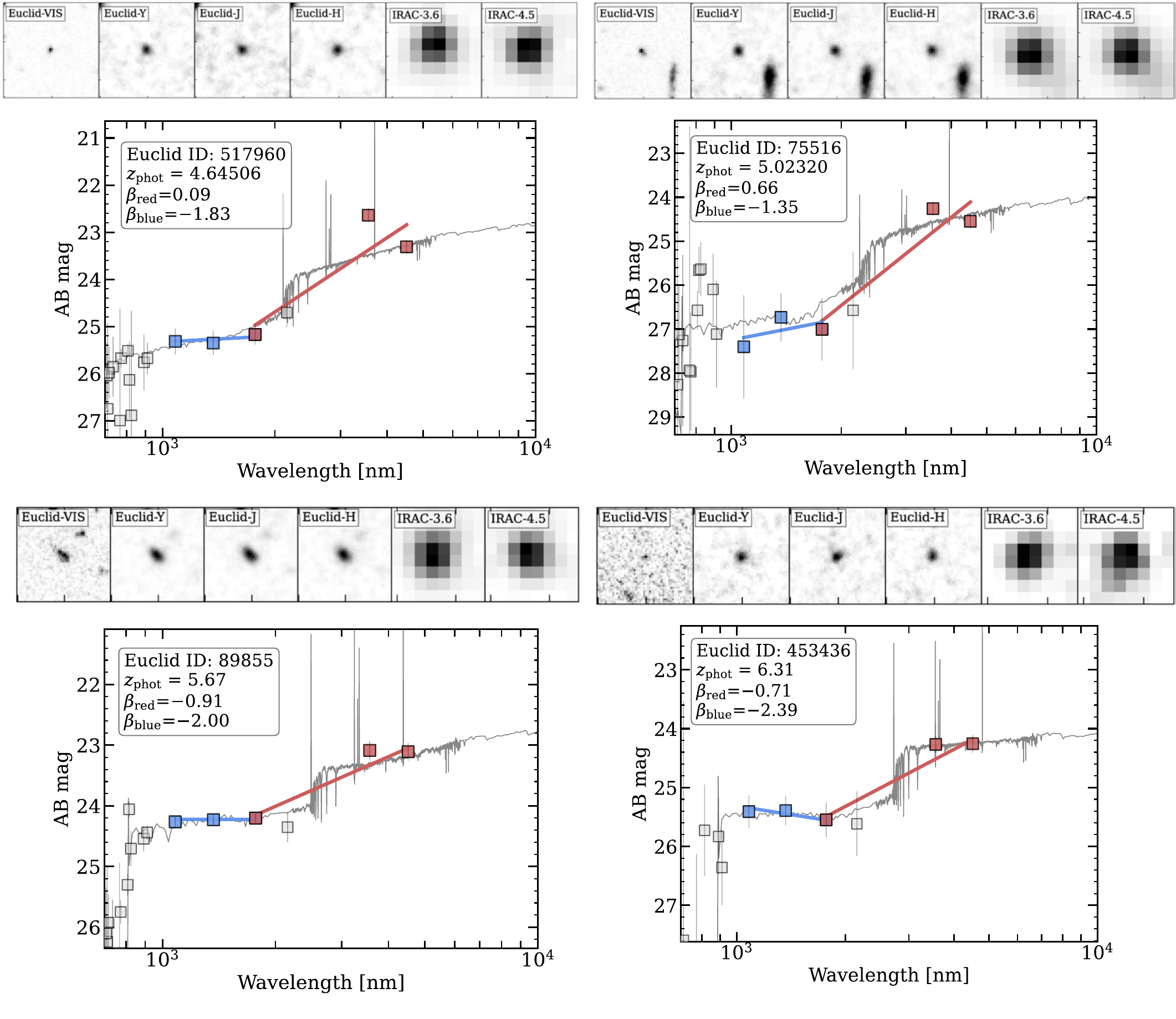}
    \caption{Example SEDs for the "Cut 1" (\textit{top}) and "Cut 2" (\textit{bottom}) sources, with $-2.8< \beta_{\rm blue} < 0.37$, and $\beta_{\rm red} > 0$ ("Cut 1") or $\beta_{\rm red} > -1$ ("Cut 2") respectively,  as defined in \citealt{Kocevski_2024}. The $\beta_{\rm red}$ and $\beta_{\rm blue}$ slopes are displayed by the red solid line and blue solid line respectively. The photometric datapoints used to measure these slopes have colours matching the corresponding continuum slope. The grey line displays the best-fit SED model. Above each spectrum are the $5''\times5''$ postage stamps, showing the bands used in our sample selection, i.e. \Euclid $\IE$, $\YE$, $\JE$,$\HE$, IRAC1, and IRAC2. 
    }
    \label{fig:spectra_stamps}
\end{figure*}

As we will discuss in  Sect.~\ref{betared_AGN}, one of the main features of interest in the double power-law sources that we aim to investigate is whether the presence of an AGN can be inferred from their SEDs.  Since \texttt{LePHARE} is by default not optimized for AGN SED modelling, we use \texttt{CIGALE} v2025.0 \citep{Boquien_2019,Yang_2020,Yang_2022} instead. 

Following the method detailed in \cite{Yang_2023}, we use the \texttt{sfhdelayed} option for the SFH, which is a standard delayed-$\tau$ module, with $\tau \in [1,5] \, \rm Gyr$ and stellar age  $\in [0.1,5] \, \rm Gyr$. For the stellar population models, to be consistent with \texttt{LePHARE}, we use the \cite{Bruzual_Charlot_2003}  models, assuming a \cite{Chabrier_2003} initial mass function (IMF) and two metallicities: one solar ($Z = Z_{\odot}$) and one sub-solar ($Z = 0.2 \, Z_{\odot}$). For the emission lines in the HII regions, we use the \texttt{nebular} module. As in \texttt{LePHARE}, we use the Calzetti--Leitherer dust attenuation law \citep{Calzetti_2000,Leitherer_2002}, allowing for $E(B-V) \in [0,1]$. 

We use the \texttt{dl2014} module for Galactic dust emission \citep{Draine_2014}. As in \cite{Yang_2023}, we allow three different values ($0.47$, $2.5$, and $7.32$) for the mass fraction of polycyclic aromatic hydrocarbon (PAH) over total dust, and the ionisation parameter $U_{\rm min}$ to vary between four possible values: 0.1, 1.0, 10, and 50. To model the AGN component, we use the \texttt{skirtor2016} model, which is in turn based on the \cite{Stalevski_2012} and \cite{Stalevski_2016} accretion disc/clumpy torus models. The $f_{\rm AGN}$ parameter, which describes the relative dominance of the AGN compared to its host galaxy, is allowed to vary between 0 and 0.99, with the $\lambda_{\rm AGN}$ parameter determining the wavelength range at which the AGN fraction, $f_{\rm AGN}$, is computed. We set $\lambda_{\rm AGN}$ to cover 0.1--1.0~$\micron$, where the accretion-disc emission is primarily found. We also set up a secondary run similar to \cite{Yang_2023} with $\lambda_{\rm AGN}$ spanning 3.0--30~$\micron$, designed to look for Type-2 AGN. This second run requires extrapolating the galaxy/AGN models, for which we do not have constraints from the photometric data. Unlike \cite{Yang_2023}, we are not only looking for Type 2 AGN, so we allow the viewing angle $i$ to vary between 0$^\circ$ and 90$^\circ$. 

Since the emission from the AGN accretion disc covers the rest-frame wavelength range $\lambda_{\rm em.} \in [0.1,10] \,\rm \micron$ \citep{Stalevski_2016},
we need observations at wavelengths longer than those corresponding to the IRAC1 and IRAC2. Although the COSMOS2020 catalogue also includes photometry from IRAC3 and IRAC4 (5.8 and 8~$\micron$, respectively), the coverage from those channels is much shallower.  Thus, out of our 6451 sources at $z>4$, only 108 have detections in either IRAC ch3 or ch4 with S/N greater than 2. We perform our \texttt{CIGALE} runs only on these 108 sources with available ch3 and ch4 photometry. As a result, the AGN-dominated sources that we find with \texttt{CIGALE} will only be a lower limit of the actual number of sources with AGN. This is because at $z>4$ a power-law component will only be evident if the  hot dusty torus surrounding the accretion disc is well exposed to the line of sight.  

In the case that these sources have a $z_{\rm spec}$ value,  we fix the redshift to that. Otherwise, we fix the redshift to the measured $z_{\rm phot}$ from our \texttt{LePHARE} runs.

\section{Sample selection} \label{sec:sample}

 Our goal here is to study $z>4$ sources which have V-shaped SEDs, with a combination of blue colours at rest-frame UV wavelengths and red colours at rest-frame optical wavelengths, and at a second stage (Sect. \ref{sec:compactness}) introduce a compactness cut to select LRD candidates. With JWST photometry, the red colour has been typically quantified using the difference of two filters ranging from F277W to F444W, and the blue colour with F115W to F200W \citep{Labbe_2023a,Kokorev_2024, Barro2026}. The intention is to encompass the light of high-$z$ objects bluewards and redwards of the Balmer and $4000 \, \rm \AA$ break. Instead  of doing this, here we followed the alternative method outlined in \cite{Kocevski_2024}, which selects sources based on the rest-frame UV and optical continuum slopes. Each slope is obtained by fitting the photometry in two or three photometric bands covering the SED. In the case of two bands, this method is identical to a colour selection.

With our \Euclid-selected \ photometric catalogue, we have measured the SED slopes using the 3 NIR \Euclid bands $\YE$, $\JE$, and $\HE$, and the \Spitzer IRAC channels 1 and 2, as they most closely approximate the range of wavelengths spanned by F115W to F444W. We define the blue filters to be $\YE$, $\JE$, and $\HE$, and the red filters to be $\HE$, IRAC1, and IRAC2. A key difference between our filter coverage and that of the traditional JWST LRDs is that we lack a filter corresponding to the JWST F277W. Even though we have  $K_{\rm s}$-band photometry in our catalogue, we chose to not use this filter because deep $K_{\rm s}$ band data is in general not available in the Euclid Deep Fields. Thus, we restricted our selection criteria to the \Euclid-\Spitzer filters. We discuss the  main  implications of the filter-set differences in  Sect.~\ref{sec:disc}.

To measure the continuum slope, $\beta$, defined such that $f_\lambda \propto \lambda^\beta$, we fitted the photometry in the red and blue bands with a linear relation described by \cite{Kocevski_2024},

\begin{equation}
\label{slopedef}
m_i =-2.5 \, (\beta +2)  \logten \left(\frac{\lambda_i}{\lambda_{\rm break}} \right) + c \; ,
\end{equation}

\noindent where $m_i$ are the AB magnitudes measured in each filter, $\lambda_i$ are the central wavelengths of the respective filters, and $\lambda_{\rm break}$ corresponding to the wavelength of the break of the V-shaped continuum. The output of these fittings gave us the rest-frame UV and optical spectral slopes, which we called $\beta_{\rm blue}$ and $\beta_{\rm red}$ respectively. 

To select the $z>4$ sources that exhibit blue UV and redder optical continuum slopes, we considered two selection cuts: 

\bigbreak 

\noindent "`Cut 1'":
\begin{equation}\label{cut1}
    \;\beta_{\rm red} > 0\;; \; -2.8 < \;\beta_{\rm blue}<-0.37\;,\\
\end{equation}


\bigbreak 

\noindent and "`Cut 2'":
\begin{equation}\label{cut2}
    \;\beta_{\rm red} > -1\;; \; -2.8 < \;\beta_{\rm blue}<-0.37\;; \beta_{\rm red} > \beta_{\rm blue}, \\
\end{equation}


The first cut corresponds to the same $\beta_{\rm blue}$ and $\beta_{\rm red}$  criteria defined in \cite{Kocevski_2024} for LRD selection. Here, however,  $\beta_{\rm blue}$ is measured from the filters $\YE$, $\JE$, $\HE$, and $\beta_{\rm red}$ is measured from the filters $\HE$, IRAC1, and IRAC2.  

The second is a less strict cut that allows for the selection of a larger source sample. In this cut, we also impose the criterion $\beta_{\rm red}>\beta_{\rm blue}$, to ensure that the SED remains V-shaped. Note that our "Cut 2" is more restrictive than the photometric criteria applied by \citet{Brazzini2026} to define LBDs.

\begin{figure}{}
    \centering
    \includegraphics[width=\columnwidth]{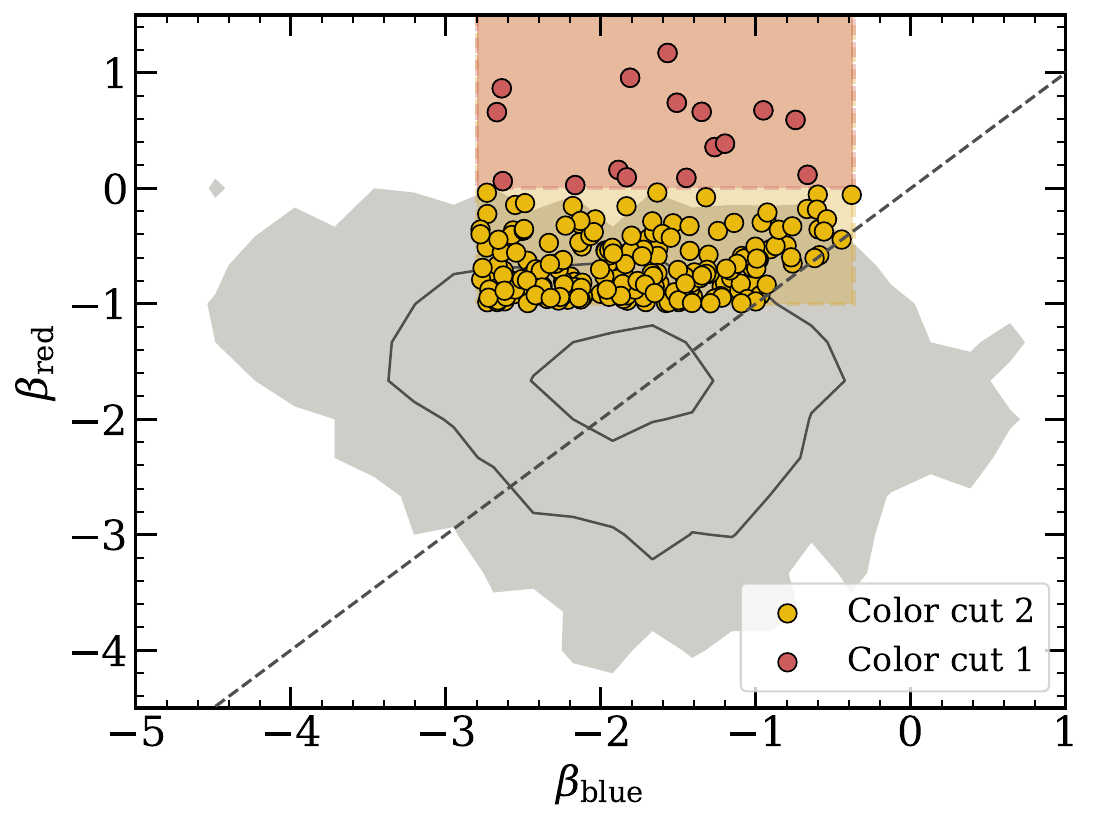}
    \caption{Continuum-slope diagram with the grey-shaded areas representing the distribution of the $z>4$ sources across the $\beta_{\rm red}$--$\beta_{\rm blue}$ plane. The shaded red and yellow regions correspond to the bounds of the colour selections for "Cut 1" and "Cut 2", as shown in Eqs.~(\ref{cut1}) and (\ref{cut2}), which select 0.25\% and 3.6\% of the $z>4$ sources, respectively. The dashed straight line corresponds to the $\beta_{\rm red}>\beta_{\rm blue}$ criterion. The red dots denote the "Cut 1" sources, and the yellow dots are the "Cut 2" sources.}
    
    \label{fig:color_betasources}
\end{figure}

 The criterion $\beta_{\rm blue}>-2.8$ in both "Cut 1" and "Cut 2" is adopted to remove brown dwarfs \citep{Kocevski_2024}. By definition, the "Cut 2" selection will include all objects in the "Cut 1". Herewith, when we refer to the "Cut 2" sources, we imply that the "Cut 1" sources are included as well. Initially, the first cut yielded 42 sources, and the second cut yielded 611 sources. However, to ensure that these colours were reliable, we performed a rigorous visual inspection of the images, particularly in the IRAC channels. Because of the low pixel resolution in the IRAC images compared to the \Euclid images, regions with multiple faint, distinct sources are highly confused in IRAC. We discarded these cases where there are multiple optical sources within the IRAC FWHM, which make for more than half of our selected sources. This step, though necessary to ensure uncontaminated photometry, could exclude potential LRD/LBD candidates. Our final clean sample contains 16 sources in the "Cut 1", and 233 sources in the "Cut 2". We show examples of the sources selected by our colour criteria in Fig. \ref{fig:color_betasources}.

At a first stage, we do not apply any compactness cut as usual for LRD selection, but study the V-shaped SED source properties as a function of compactness instead (see Sect. \ref{sec:compactness}). We define compactness, $C_{\HE}$, as the ratio between the fluxes in the $\ang{; ;0.7}$-diametre aperture and the $\ang{; ;1.0}$-diametre aperture in the \Euclid $\HE$ band. We selected the $\ang{; ;0.7}$-diametre aperture as the inner radius, as it is roughly twice the PSF FWHM of the $\HE$ band. For the “outer radius”, having tested a range of apertures, we found that using the $\ang{; ;1.0}$-diametre aperture most effectively selected the sources that we identified as being compact by visual inspection.  We give more details about our compactness criterion in Sect.~\ref{sec:compactness}.

\section{Properties of the \Euclid/IRAC V-shaped-SED sources at $z>4$}\label{sec:dplprop}

\subsection{Redshift distribution}\label{dpl_redshift}

\begin{figure}
    \centering
    \includegraphics[width=\columnwidth]{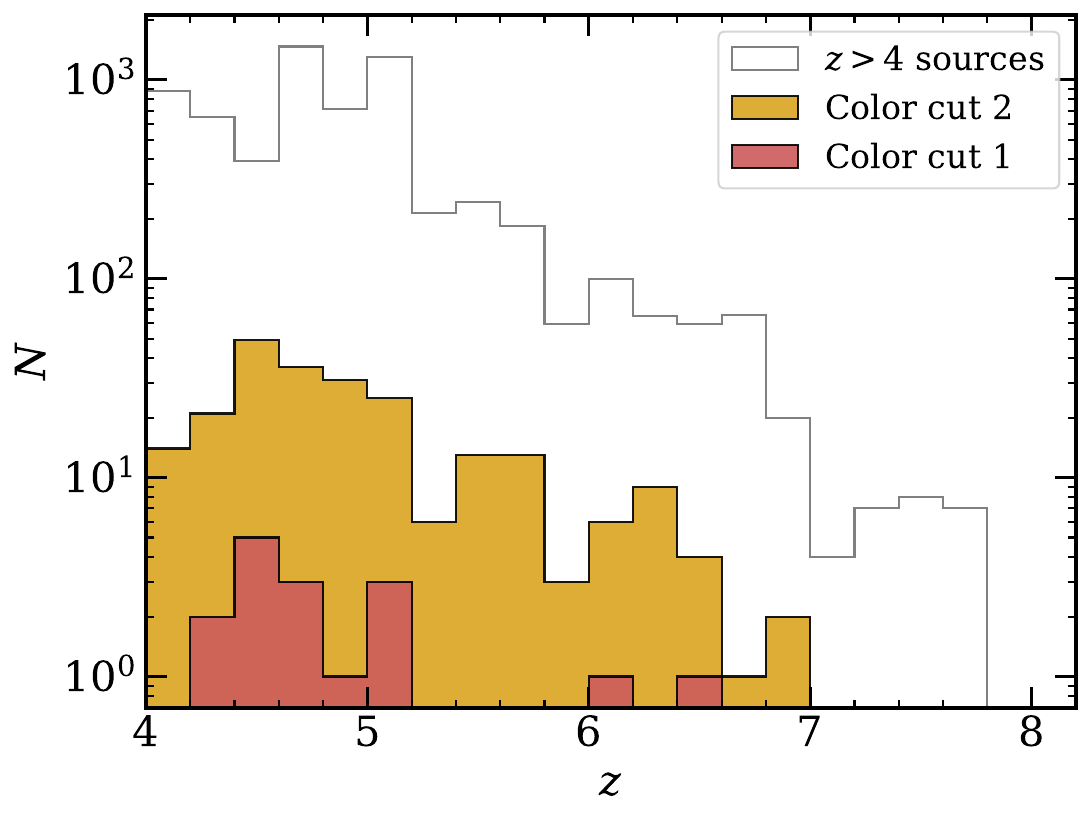}
    \caption{Redshift distribution  of $z>4$ sources (empty histogram), with the "Cut 1" and "Cut 2" sources in red and yellow respectively. }
    \label{fig:redshift_dist}
\end{figure}

\begin{figure*}
    \centering
    \includegraphics[width=\textwidth]{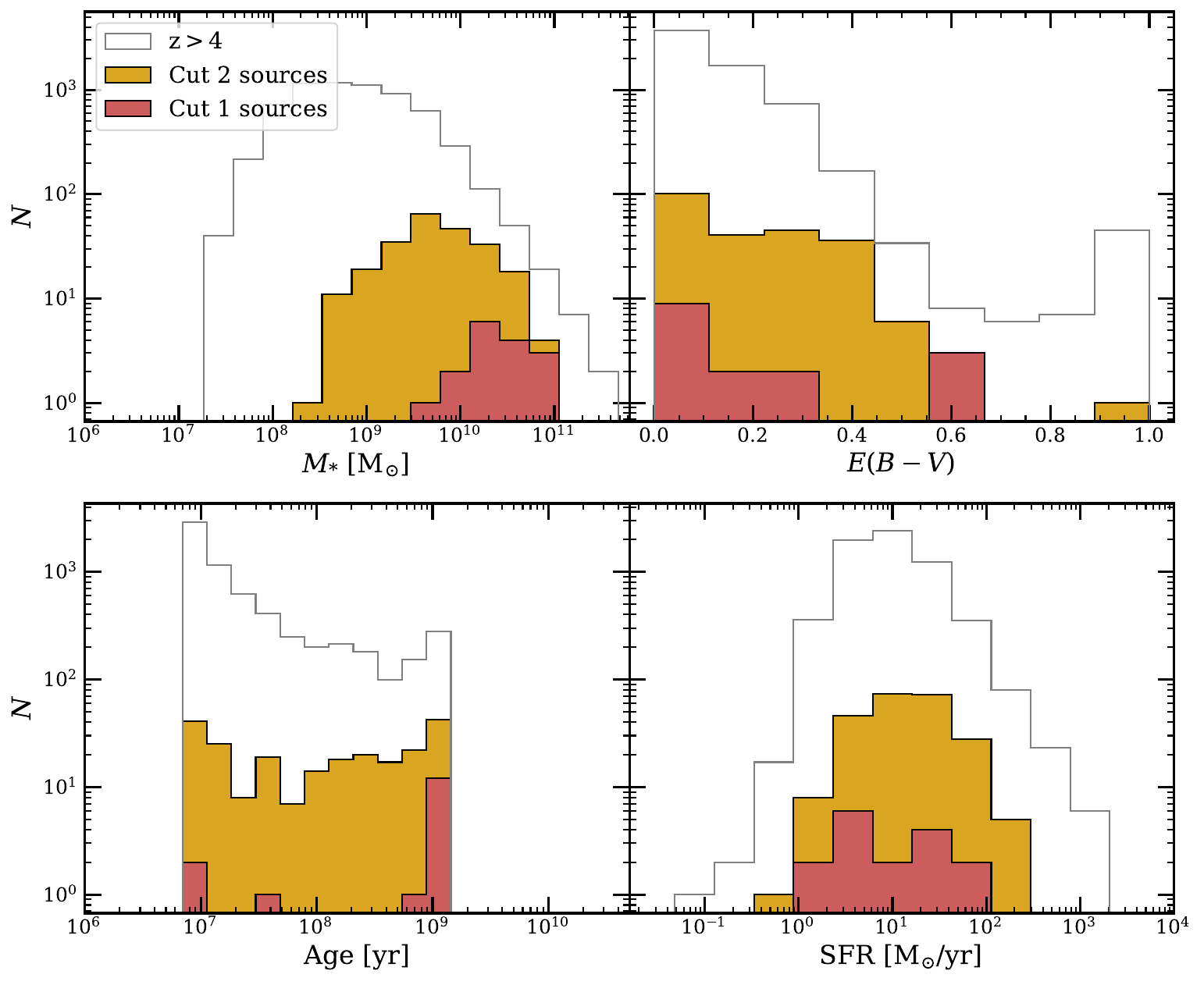}
    \caption{Distributions of our two double power-law sources sample populations (red for "Cut 1" and yellow for "Cut 2"), compared to all $z>4$ sources (empty histogram). We show the distributions for stellar mass, colour excess $E(B-V)$, SFR, and age, starting from the top left and clockwise. All SFR values shown in this figure have been calculated from the rest-frame UV luminosities, assuming that all the UV light is produced by star formation, with negligible AGN contribution. }
    \label{fig:betaproperties}
\end{figure*}

The redshift distribution for $z>4$ sources, as obtained from \texttt{LePHARE} and supplemented with $z_{\rm spec}$ from \cite{Khostovan_2025}, is shown in Fig. ~\ref{fig:redshift_dist}.  We find that only a small fraction of our high-redshift sources exhibit  double power-law SEDs, with 0.25\% and 3.6\% of the $z>4$ sources falling under the criteria for "Cut 1" and "Cut 2" respectively.

Our double power-law selected galaxy samples contain a much higher percentage of $z<4$ sources than JWST-selected LRD samples, which almost exclusively consist of $z>4$ sources \citep{Kokorev_2024,Kocevski_2024, Barro2026}. This is because our sources are brighter and also partly because of the different filters used in the selection (albeit there is a large overlap between the sources selected with V-shaped SEDs using $\HE$ or $K_{\rm s}$ as pivotal passbands, see Sect.~\ref{sec:diff}). We find that our photometric selection contains sources at all redshifts, with the stricter criterion in "Cut 1"  containing a $z>4$ source percentage of 8.9\% compared to  7.3\% in "Cut 2".

In any case, here we only focus on the $z>4$ sources in our double power-law selected sample, as they allow for a direct comparison with JWST-selected LRDs/LBDs. From Fig.~\ref{fig:redshift_dist} we can see that our $z>4$ sources in "Cut 2" follow closely the general redshift distribution of all $z>4$ galaxies, except that we find virtually no source at $z>7$. Part of the reason is because of our $\beta_{\rm blue}$ cut, which excludes red sources at wavelengths where the Lyman-break is expected to be observed. Another reason is that some of the $z>7$ sources have photometry and best-fit spectra with flat continua above 1~$\micron$, thereby being excluded from our double power-law selection. We also note from previous (JWST) studies that the visibility of V-shaped SED sources drops significantly at $z>7$ \citep{Rinaldi_2024,Pacucci_2025,Billand_2025}. 

\subsection{Distribution of physical properties}\label{sec:dpl_physprop}

Figure~\ref{fig:betaproperties} shows the property distribution of the double power-law galaxies with "Cuts 1" and "2" at $z>4$, compared to those of all galaxies at the same redshifts. We study stellar mass, SFR, colour excess $E(B-V)$, and best-fit age. The SFR have been inferred based on the rest-frame UV luminosities (calculated from the observed photometry).

We see that the stellar masses for the "Cut 1" sources are mostly contained within $[\rm 10^{10},10^{11}] \, \it M_{\odot}$, with a median $\rm \logten(\it{M_\ast }/\it M_\odot) \rm  = 10.36^{+0.30}_{-0.32}$ indicating that this selection is comprised of relatively massive galaxies at $z>4$. On the other hand,  the sources in "Cut 2" cover a wider stellar mass range between $10^{8} \, \it M_\odot$ and  $10^{11} \, \it M_{\odot}$, with a median $\rm \logten(\it{M_\ast} /\it M_\odot) \rm = 9.74^{+0.49}_{-0.51}$. This implies that a larger variety of sources are being included in this more relaxed cut. In any case, the median stellar masses for both cuts of our double power-law sources are higher than that of all $z>4$ sources, which is $\rm \logten(\it{M_*}/\it M_\odot) \rm  = 8.82^{+0.67}_{-0.58}$. 

With respect to the dust extinction, we find less variations between the medians of the double power-law sources and the general $z>4$ galaxy population. All their median values of the colour excess are consistent within the 68\% intervals:  $E(B-V) = 0.1^{+0.31}_{-0.14}$ and  $E(B-V) = 0.2 ^{+0.16}_{-0.17}$, for "Cut 1" and "Cut 2" sources, respectively. And for the general $z>4$ galaxy sample we have  $E(B-V) = 0.1^{+0.17}_{-0.09}$. This means that, while highly reddened sources are found at $z>4$ as expected \citep[e.g.,][]{Caputi_2015}, the double power-law sources do not preferentially select them.  This is direct consequence of the $\beta_{\rm blue}$ cut.

Nonetheless, we find a small group of three sources with $E(B-V)=0.6$ among the "Cut 1" double power-law sources. These are very dusty sources with intermediate stellar masses $M_\ast \in  [10^{9.7}, 10^{10.8}] \, \it M_\odot$. Notably, they are  highly star-forming, with $\rm SFR \in [30,60] \, \it M_\odot \rm \, yr^{-1}$, and very young, with best-fit ages lower than $10^{7.5}$ yr. In addition, they are slightly more compact than the median of all $z>4$, with compactness values between 0.72 and 0.75  (while $C_{\rm median \,\it z \rm >4}$=0.70; see Sect.~\ref{sec:compactness}). In summary, these are compact, very dusty galaxies that are intensively forming new stars. Note that, albeit compact,  these sources do not satisfy the compactness criterion that we impose later to select LRD/LBD candidates.

The best-fit age distribution of the $z>4$ galaxies appears to be bimodal with peaks at $10^7$ and $10^9$ years, and with some preference for younger ages in general. Interestingly, however, the \textit{Cut~1} V-shaped-SED sources have a preference for older ages instead, with $\sim 10^9$~yr for 81\% of them. 

To evaluate the SFR of our sources, we measured the luminosity from the flux in the filter encompassing rest-frame  $\lambda_{\rm rest} = 2000 \, \rm \text{\AA}$. We also applied dust-extinction corrections using the \cite{Calzetti_2000} reddening law, making use of the best-fit output $E(B-V)$ values from the \texttt{LePHARE} SED-fitting results. We then converted the dust-corrected fluxes into monochromatic luminosity $L_{2000\, \text{\AA}}$, and then used the relation derived for a Salpeter IMF by \cite{Kennicutt_1998},

\begin{equation}
\rm SFR_{2000 \, \text{\AA}}\,/\,(\it M_\odot \,\rm  yr^{-1}) = 1.4 \times 10^{-28} \, \mathit{L}_{2000 \, \text{\AA}} \,/\, (erg~ s^{-1} Hz^{-1}) \; ,
\end{equation}

\noindent to which we multiplied by a factor 0.63 to convert into a Chabrier IMF \citep{Madau_Dickinson_2014}. We then applied metallicity-dependent calibrations by \cite{Theios_2019} for our two metallicity populations.

In the bottom right plot of Fig. \ref{fig:betaproperties} we show the resulting SFR distribution. For the $z>4$ sources with stellar masses similar to the "Cut 2" sources ($M_{\ast} \rm  \gsim \, 10^{8.5} \it \, M_{\odot}$), we find a median $\logten(\rm SFR/\it \, M_{\odot} \rm \, yr^{-1}) = 1.09^{+0.46}_{-0.41}$. The SFR medians of the V-shaped-SED source distributions are consistent within the error bars:  $\logten(\rm SFR/\it \, M_{\odot} \rm \, yr^{-1}) = 0.76^{+0.72}_{-0.41}$ and  $\logten(\rm SFR/\it \, M_{\odot} \rm \, yr^{-1}) = 1.13^{+0.47}_{-0.46}$ for the "Cut 1" and "Cut 2" sources, respectively. So no SFR enhancement appears to be evident amongst the V-shaped SED sources.

\begin{figure}
    \centering
    \includegraphics[width=\columnwidth]{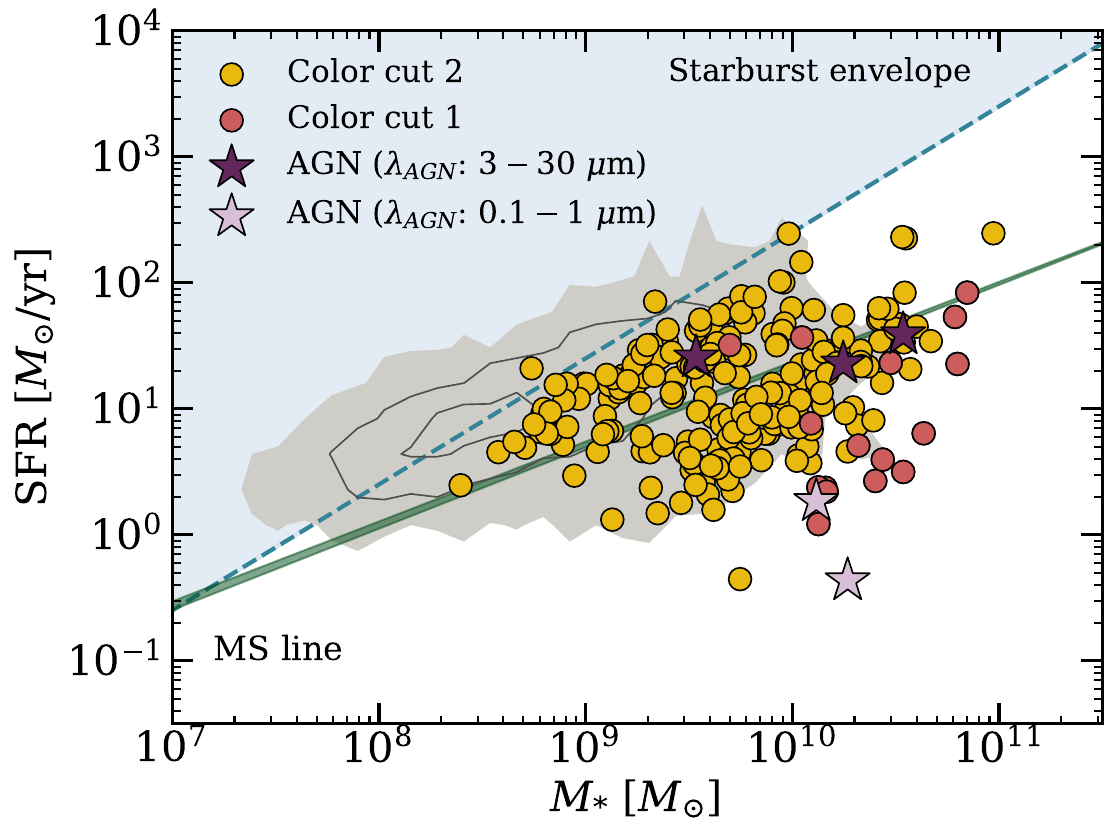}
    \caption{Stellar mass versus SFR, with the shaded grey regions representing the distribution of all $z>4$ sources, with the contours highlighting the 84th and 95th percentiles. The starburst envelope, as defined in \cite{Caputi_2017} and  \cite{Caputi_2021}, is marked by the shaded blue region, above the dashed blue line. The median of the star-formation main sequence (MS) for $z \in [4,6]$  \citep{Rinaldi_2022} is shown with a green line. The coloured dots denote the sources in "Cut 1" (red) and "Cut 2" (yellow), and the stars mark where the \texttt{CIGALE}-identified AGN appear on the $\it M_*$--SFR plane. Here the SFR of these identified AGN have been corrected, such that only the stellar component is considered.}
    \label{fig:mass_compactness_sfr_betared}
\end{figure}

In Fig. \ref{fig:mass_compactness_sfr_betared} we show the stellar mass versus SFR plane, with the region above the dashed line indicating the starburst regime, as empirically defined by \cite{Caputi_2017, Caputi_2021}, and the solid line indicating the median of the star-formation main sequence (MS), as fitted by \cite{Rinaldi_2022}.
We find that the majority of the "Cut 2" sources lie around the star-formation MS, below the starburst division line. There are only 3 sources ($\sim 1.3\%$) out of 233 that are within the starburst envelope, and they generally have best-fit younger ages ([$10^7$,$10^8$] yr).  In the more restrictive "Cut 1" group, we note that all of these sources, which are mainly old and massive, are around or below the star-formation MS. 

In Fig. \ref{fig:mass_compactness_sfr_betared}, we also plot the 5 \texttt{CIGALE}-identified AGN on the $M_*$--SFR plane, which we will discuss in further detail in Sect.~\ref{betared_AGN}. We emphasize that, since we measure SFR from the $2000 \,\text{\AA}$ luminosity, if there is an AGN component, the SFR will be overestimated. In light of this, we have corrected the $\rm SFR_{2000 \,  \text{\AA}}$ of these CIGALE-identified AGN by only taking into account the contribution of the stellar models at $2000 \,  \text{\AA}$ as computed by \texttt{CIGALE}. This reduces the SFR values of the \texttt{CIGALE}-identified AGN, bringing them down to around the MS line, which more accurately represents the property of their host galaxy. 


\subsection{Compactness distribution of the double power-law sources}
\label{sec:compactness}

\begin{figure*}
    \centering
    \includegraphics[width=\textwidth]{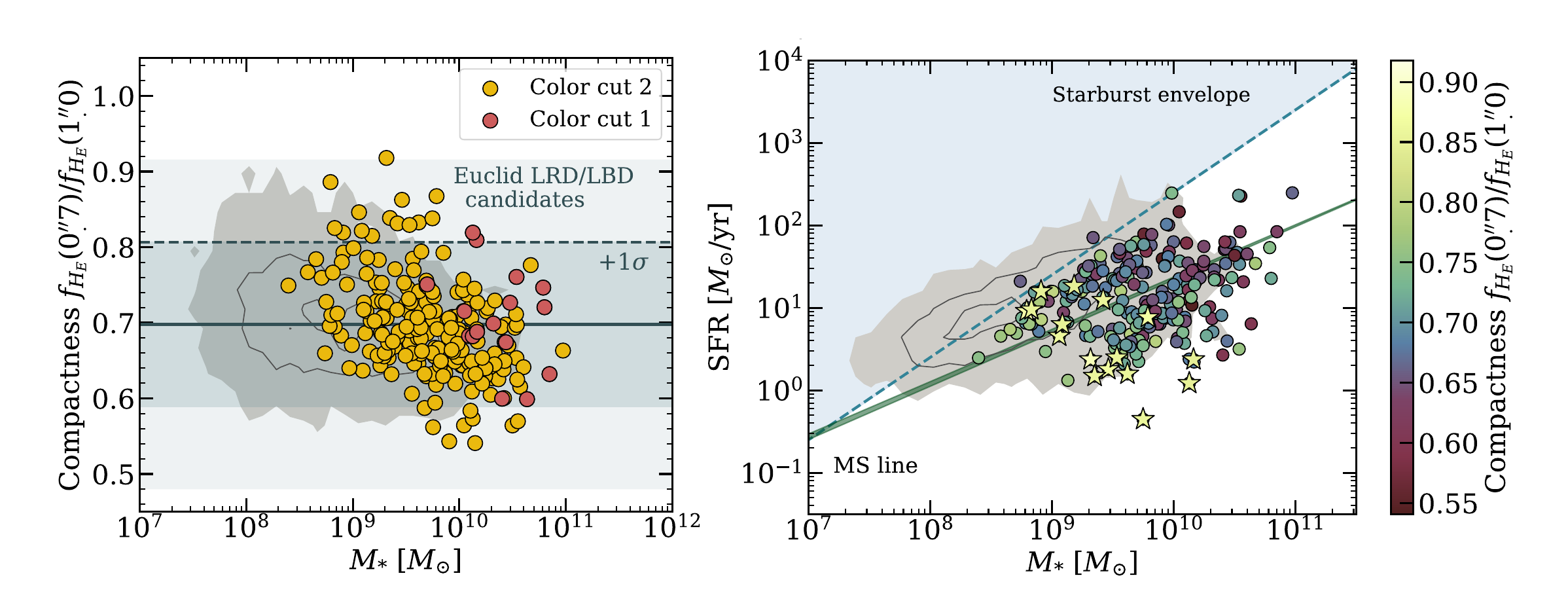}
    \caption{{\em Left:} Stellar mass versus compactness, defined as $ f_{\HE}$($\rm \ang{; ;0.7}$)/$f_{\HE}$($\ang{; ;1.0}$), with the shaded grey regions representing the distribution of all $z>4$ sources. The yellow dots denote the "Cut 2" sources, and the red dots are the more restrictive "Cut 1" sources. The median of the compactness distributions of the $z>4$ population is shown by the solid line, and the shaded blue areas denote the 1$\sigma$ and 2$\sigma$ regions. All V-shaped SED sources above the $+1\sigma$ line are considered to be the \Euclid LRD/LBD candidates. {\em Right:} Stellar mass versus SFR, like before, with the shaded grey regions representing the distribution of all $z>4$ sources. The dots represent the double power-law sources, colour-coded by their compactness. The 16 LRD/LBD candidates are marked with stars. }
    \label{fig:mass_compactness}
\end{figure*}

As one of the defining characteristics of LRDs/LBDs is their compact morphology at rest-frame optical wavelengths, we compare the degree of compactness of our V-shaped SED sources to the rest of our high-redshift galaxy sample. We measure the source compactness, $C_{\HE}$, based on the \Euclid \HE-band images, considering the flux-density ratio $ f_{\HE}$($\rm \ang{; ;0.7}$)/$f_{\HE}$($\ang{; ;1.0}$). As explained before, the inner aperture diametre roughly corresponds to twice the PSF FWHM in the $H_E$, and the outer aperture diametre has been empirically chosen to maximize the selection of sources that appear to be compact by visual inspection.

The median compactness values for the three populations, namely all $z>4$ galaxies, "Cut 2" double power-law sources, and "Cut 1" double power-law sources, are 0.70, 0.69, and 0.71 respectively. To quantify whether we consider a source to be compact or not, we refer to the median compactness of all $z>4$ sources. We define any source $>\,1\sigma$ above the median to be `compact', i.e., with $C_{\HE} > C_{\rm median \, \it  z \rm >4} +$1$\sigma$ $=$ 0.81 . To put this terminology in a more general context, one has to consider that $z>4$ galaxies are on average more compact than lower redshift galaxies, so our compact definition here is significantly strict.

In the left panel of Fig. \ref{fig:mass_compactness}, we show the distribution of compactness against stellar mass of our V-shaped SED sources, with the 1$\sigma$ and $2\sigma$ regions indicated. For our "Cut 2" sources, only 0.43\% are classified as very compact (2$\sigma$ above the median of all $z>4$), and 6.9\% are compact, while 3.9\% of them are extended, and none is very extended (2$\sigma$ below the median of all $z>4$). We find that 99.6\% of the "Cut 2" sources fall within $\pm2\sigma$ around the median of all $z>4$ sources.  The "Cut 1" double power-law sources are less heterogeneous in compactness distribution. None of them are classified as extended, as all their compactnesses are higher than 1$\sigma$ below the median. We find that 12.5\% of them are compact, but none are extremely compact.

From Fig. \ref{fig:mass_compactness}, we see that the more massive sources tend to be more extended, so imposing a compactness criterion from the beginning would exclude the most massive double power-law SED sources. We find the compact sources below the starburst division line on the lower-mass end of the star-formation MS (right panel of Fig. \ref{fig:mass_compactness}). The more extended sources, on the other hand, though still mainly below the starburst envelope, are generally more star-forming on the higher-mass end.  

LRDs/LBDs have by definition compact morphologies. So we will use our compactness criteria, along with the colour cuts defining the V-shaped SEDs, to identify LRD/LBD candidates amongst our \Euclid-selected $z>4$ galaxies. Particularly, there are 16  V-shaped SED sources that we classified as compact, as they have compactness greater than 0.81, i.e., $1\sigma$ above the median of all $z>4$ galaxies.  We consider these 16 to be  our most secure LRD/LBD candidates (and we will simply call them \Euclid LRDs/LBDs hereafter). Out of these 16 compact, V-shaped SED sources that satisfy "Cut 2", only two of them satisfy the stricter "Cut 1". 

The list of these  \Euclid LRDs is presented in Table~\ref{table:lrd_candidates}. These compact sources span the stellar-mass range  $[\rm 10^{8.5},10^{10.5}] \, \it M_{\odot}$. They have relatively lower SFRs than the rest of the double power-law SED sources, spanning  $[\rm 10^{-0.5},10^{1.5}] \, \it M_{\odot}\rm \, yr^{-1}$. As a consequence, about a half of them lie significantly below the star-formation main sequence (MS). Remarkably, half of the \Euclid LRDs have maximal best-fit ages, i.e., they are basically as old as the Universe at their redshifts. This is in contrast with the general V-shaped SED galaxy population, whose best-fit ages span a wide range of values. 

Although the \Euclid LRDs/LBDs constitute a small percentage ($\sim 7\%$) of all V-shaped SED sources, they comprise $\sim 15\%$ of all those with low stellar masses ($\lsim 10^{9.48} \, \it M_\odot$). This is likely related to the galaxies becoming more extended as its stellar mass increases, and thus falling out of the compactness criterion that defines the LRD selection.  

\label{LRD_candidates}
\begin{table}
\begin{minipage}{\columnwidth}
  \centering
\scriptsize
\caption{List of our 16 \Euclid LRDs/LBDs, i.e the $z>4$ sources fulfilling the colour criteria described in Section~3 and the compactness criterion ($C_{\HE} > C_{\rm median \, \it  z \rm >4} + 1\sigma$), where $C_{\rm median \, \it  z \rm >4}$ is the median compactness of all $z>4$ sources.}
\label{table:lrd_candidates}
\begin{tabular}{lcccccc}

\\ \hline \hline \\ ID & RA & Dec & $ z_{\rm phot}$ & $\beta_{\rm red}$ & $\beta_{\rm blue}$ & $\rm log_{10}\it(M_{\ast}/  M_{\odot})$ \smallskip

\\ \cline{1-7}

18218 & 149.92392 & 1.99088 & 4.66 & $-0.82$ & $-1.19$ & 9.18\\ 
54295 & 149.96968 & 1.94516 & 4.09 & $-0.96$ & $-2.21$ & 9.06\\ 
75516 & 150.11720 & 2.05305 & 5.02 & $0.66$ & $-1.35$ & 10.16\\ 
199013 & 149.95035 & 2.27144 & 4.68 & $0.95$ & $-1.81$ & 10.13\\ 
390377 & 149.97565 & 2.45276 & 4.58 & $-0.32$ & $-0.76$ & 9.62\\ 
595447 & 149.88176 & 2.70738 & 5.15 & $-0.28$ & $-2.13$ &8.79\\ 
67788 & 149.95831 & 2.01873 & 4.94 & $-0.96$ & $-2.35$ & 8.91\\ 
101791 & 150.27891 & 1.91332 & 4.82 & $-0.86$ & $-2.67$ & 9.32\\ 
227356 & 150.04432 & 2.11737 & 4.58 & $-0.41$ & $-2.57$ & 9.46\\ 
287547 & 150.07482 & 2.38230 & 5.13 & $-0.80$ & $-2.56$ & 8.83\\ 
299459 & 150.26879 & 2.15384 & 4.45 & $-0.04$ & $-2.73$ & 9.35\\ 
328458 & 150.22995 & 2.28580 & 5.02 & $-0.14$ & $-2.55$ & 9.75\\ 
395286 & 149.86713 & 2.47838 & 4.44 & $-0.36$ & $-2.77$ & 9.79\\ 
454704 & 150.05552 & 2.49806 & 4.69 & $-0.35$ & $-2.50$ & 9.42\\
521038 & 150.43772 & 2.50521 & 4.93 & $-0.69$ & $-2.76$ & 9.08\\
606474 & 149.99299 & 2.77905 & 4.48 & $-0.40$ & $-2.78$ & 9.53\\
    \hline  
\end{tabular}

\end{minipage}
\end{table}

\subsection{AGN identification amongst the V-shaped-SED sources at $z>4$}\label{betared_AGN}
\label{sec:agnidentif}

A proper identification of BLAGN amongst our V-shaped-SED sources requires dedicated spectroscopy. 
In the absence of it, we apply a suite of other methods to try to infer an AGN presence. First, we cross-matched our \textit{Cut~2}, which includes \textit{Cut~1}, V-shaped-SED sources with existing AGN catalogues, based on ancillary data across different wavelengths. In particular, the compilation catalogue by \citet{Khostovan_2025} has spectroscopically identified AGN (based on broad-line features), but we do not find any matches with our sources. Actually only a very small fraction of the sources in their catalogue are classified as BLAGN at $z\geq 4$ and none coincides with our double power-law SED sources.

We do not find any matches either with sources in the \textit{Spitzer} MIPS $24 \, \rm \micron$ catalogue \citep{Sanders_2003}, but we do find one match with the \textit{Chandra} X-ray AGN \citep{Civano_2016}, indicating that there is at least one X-ray-detected AGN within our sample. Interestingly, this X-ray AGN source is not compact, with a compactness value 0.66, which is less than the median compactness for all $z>4$ sources. 

We also cross-matched our sources with the recent catalogue of $z\geq6$ `quasar' candidates from \cite{Andika_2025} and found 4 matched objects. The AGNs identified from \cite{Andika_2025} come from the COSMOS-Web field with photometry up to and including the MIRI F770W filter. These AGN are also not very compact, with only 1 of them having a compactness value that is greater than the median compactness of the $z>4$ sources. It is worth noting that although these sources are not considered compact according to our definition, they are also not extended. 

The \cite{Andika_2025} AGN are identified via SED fitting with \texttt{CIGALE} based on a similar photometric coverage to our sources, but imposing an AGN fraction $\geq$ 0.2. This is a more lenient AGN fraction criterion than ours, which only takes sources with AGN fraction $>0.5$. Unfortunately, our matched objects do not have detections in IRAC ch3 or ch4, so it is difficult to say whether relaxing the AGN definition to match the criterion by \cite{Andika_2025} would have yielded more objects in common.

Furthermore, we checked whether any of our V-shaped SED sources would be identified as an AGN based on their \texttt{LePhare}SED fitting. In \texttt{LePhare}, we have considered empirical quasi-stellar objects (QSO) templates \citep{Polletta_2008} along with the synthetic stellar templates for SED fitting, so we checked which of our sources preferred a QSO solution (i.e. whether $\chi^2_{\rm QSO}  < \chi^2_{\rm gal.}$). Among our "Cut 1" galaxies, we find that 25\% of the sources (4 out of 16) prefer QSO models, which also lead to photometric redshift solutions $\geq$ 4. For the "Cut 2" sources, 13.3\% of them (31 out of 233) had best-fit QSO models, but only 5.2\% (12 out of 233) had resulting $z_{\rm phot} \geq 4$.  But out of these 12 sources preferring the QSO model, only 1 of them has a galaxy solution that differs by more than 1$\sigma$ from the QSO solution, indicating that the remaining 11 have QSO and galaxy model solutions that are comparable. One of our \Euclid LRDs, which is a compact V-shaped double power-law source satisfying the "Cut 2" criteria, falls in this high-redshift QSO group. 

Then, as described earlier in Sect.~\ref{sec:agn_identification}, we also used \texttt{CIGALE} to see if any of these double power-law sources had a dominant AGN presence. Unfortunately, the \textit{Spitzer} mid-IR coverage beyond $\rm 5 \, \micron$ is shallow in the COSMOS field, and only 8 out of the 233 double-power-law-SED sources are detected in IRAC ch3 and/or ch4 with S/N$\,>2$. The SED decomposition into stellar/AGN component with \texttt{CIGALE}  makes most sense in these 8 cases. We will later discuss whether using photometry only up to ch2 ($4.5 \, \rm \micron$) in these 8 cases can also lead to a similar AGN identification, such that we can extend this SED decomposition to all the double-power-law-SED sources.

For the 8 sources with detections up to IRAC ch3 and/or ch4 we find the following.  When we request \texttt{CIGALE} to compute the AGN fraction at $\lambda_{\rm AGN} \in [0.1,1] \, \micron$ in the SED, we obtain 2 out of 8 (25\%) sources with $f_{\rm AGN} >0.5$. Using the wavelength range $\lambda_{\rm AGN} \in \, [3,30] \, \micron$ to calculate $f_{\rm AGN}$ yields more AGN-dominated sources, giving us 3 out of 8 (37.5\%). There are no overlapping sources between these two groups of AGN candidates, which indicates that 62.5\% of our double power-law sources with detections in IRAC ch3 and ch4 are AGN candidates. Nonetheless, we note that the latter group is obtained from an SED extrapolation, as we have no constraints on rest-frame $\lambda_{\rm emit} \geq 1.5 \, \rm \micron $ with our photometric data, so the latter 3 additionally found AGN candidates must be considered with caution.  

We then ran \texttt{CIGALE} with photometry only up to IRAC ch2 for the sources with detections in IRAC ch3 and ch4. The intention is to test whether AGN would be similarly recognised in the absence of ch3 and ch4 data. In this case, only one of the two sources classified as AGN from the $\lambda_{\rm AGN} \in [0.1,1] \, \micron$ wavelength range was classified as AGN with this run with restricted photometry. This suggests that without IRAC ch3 and ch4, \texttt{CIGALE} is not able to reproduce the same AGN fractions as when we do have IRAC ch3/ch4 photometry. 

Interestingly, we do not find any overlap either between the \texttt{CIGALE}-classified AGN with photometry up to IRAC ch4 and the \texttt{LePHARE}-classified QSO sources, nor with the one cross-matched \textit{Chandra} X-ray AGN or the 4 AGN candidates from \cite{Andika_2025}. Combining the AGN from the literature and the QSO/AGN-classified objects from our SED-fitting gives us a total of 22 potential AGN candidates within all our V-shaped-SED sources, which we have compiled in Table \ref{table:agn_matches}.


\begin{table}
\begin{minipage}{\columnwidth}
  \centering
\scriptsize
\caption{List of identified AGN (from the literature and our own SED fitting) amongst our V-shaped-SED sources. 
}
\label{table:agn_matches}
\footnotesize
\begin{tabular}{p{1.cm} r r r r}
\toprule
\\ ID & RA & Dec & $z_{\rm phot}$ & Source   \smallskip
\\  & [J2000] & [J2000] &  &    \smallskip
\\ 
\midrule
50719 & 150.00716 & 1.91795 & 6.88 & CHANDRA  \\ 
73322 & 150.06892 & 2.04407 & 4.84 & \cite{Andika_2025}  \\ 
86258 & 150.18258 & 2.10186 & 6.22 & " \\ 
177566 & 149.93324 & 2.16685 & 6.03 & " \\ 
25139 & 150.10264 & 2.22486 & 6.38 & " \\

417492 & 149.84320 & 2.59988 & 5.56 & \texttt{CIGALE}/   \\	
455360 & 150.12668 & 2.50118 & 5.56 & $\rm \lambda_{AGN} \in [0.1,1] \, \micron$  \\ 

53626 & 150.17975 & 1.94080 & 5.39 & \texttt{CIGALE}/  \\ 
89855 & 149.97184 & 2.11818 & 5.69  &  $\rm \lambda_{AGN} \in [3.0,30] \, \micron$  \\ 
191419 & 149.77469 & 2.23431 & 4.28  & " \\ 

199013$^\ast$ & 149.95035 & 2.27144 & 4.68 & \texttt{LePhare} QSO  \\ 
247267 & 149.99375 & 2.20737 & 5.29 &  " \\ 
256310 & 150.06376 & 2.24680 & 6.07 &  " \\ 
278638 & 150.14443 & 2.34440 & 4.72 &  " \\ 
415547 & 149.86213 & 2.58777 & 4.24 &  " \\ 
478653 & 150.01298 & 2.60679 & 6.07 &  " \\ 
596626 & 149.90445 & 2.71467 & 5.15 &  " \\ 
620534 & 150.30563 & 2.63562 & 4.34 &  " \\ 
191261 & 149.75164 & 2.23344 & 5.39 &  " \\ 
215802 & 149.72876 & 2.34728 & 4.65 &  " \\ 
280728 & 150.19494 & 2.35323 & 4.66 &  " \\ 
606348 & 149.83065 & 2.77814 & 6.46 &  " \\ 

\bottomrule

\end{tabular}
\tablefoot{
\tablefoottext{$\ast$ }{Compact source, i.e., LRD/LBD}}
\end{minipage}
\end{table}

\section{Standard AGN as a mostly disjoint population to V-shaped SED sources at $z>4$}\label{sec:agn}

\begin{figure}
    \centering
    \includegraphics[width=\columnwidth]{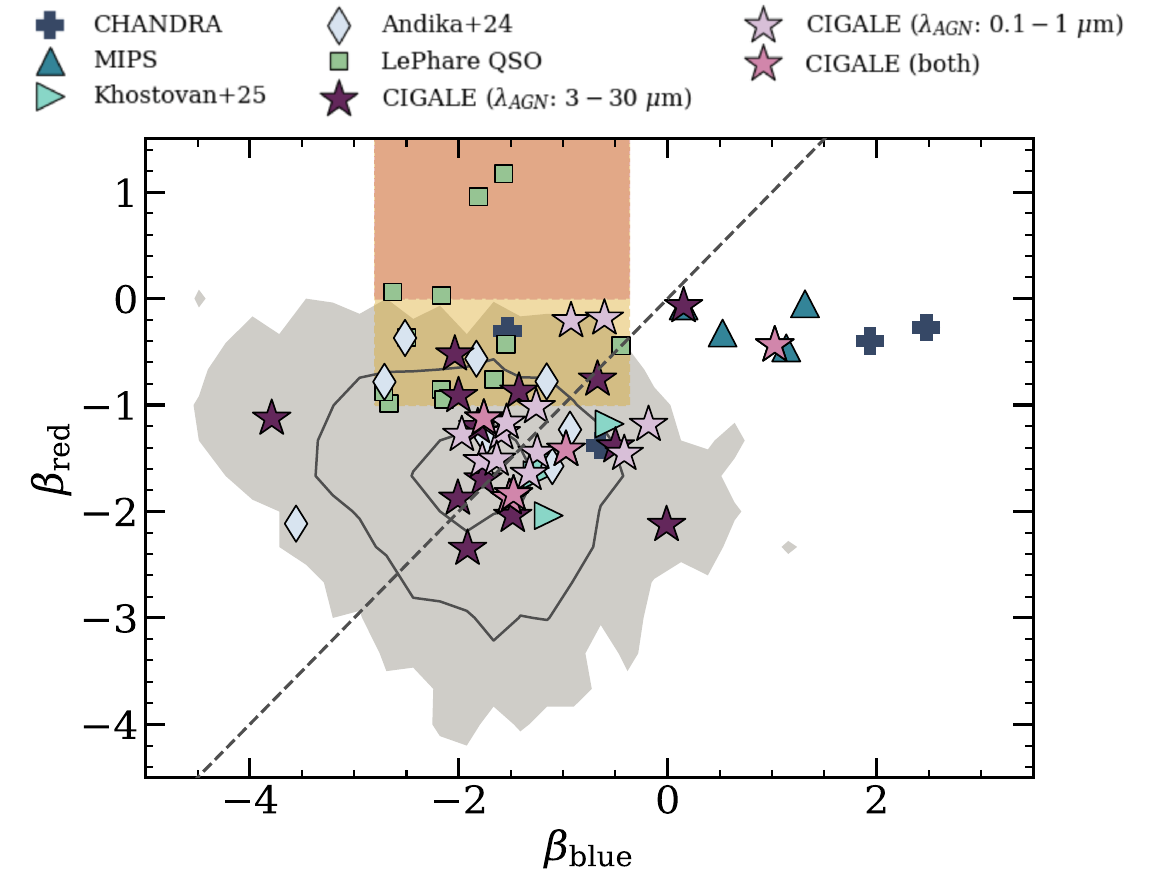}
    \caption{Classical AGN, recognised with different criteria, on the $\beta_{\rm red}$--$\beta_{\rm blue}$ diagram.   As previously, the red and yellow shaded regions represent the boundaries of the $\beta$ "Cut 1" and "Cut 2" selections.  }
    \label{fig:color_classicAGN}
\end{figure}

A number of studies have found that LRDs/LBDs have X-ray luminosities that are significantly weaker than expected in comparison to other known (`standard') AGN \citep{Yue_2024,Maiolino_2025}. This suggests that if there were LRDs/LBDs within our double power-law sample, we should expect little overlap between them and known AGNs. Given this, it is not surprising that only a small fraction of our double power-law sources have a match with the standard AGN identified in ancillary catalogues at $z>4$. To understand this issue further, we investigate where the classical AGN that  are present amongst our $z>4$ galaxies lie on the $\beta_{\rm red}$--$\beta_{\rm blue}$ diagram.  This is shown in Fig.~\ref{fig:color_classicAGN}.

From our CIGALE runs on all the $z>4$ galaxies with IRAC ch3 and ch4 detections, which amounts to 102 sources, we have 17 sources with AGN fractions $>0.5$ as measured from the $\lambda_{\rm AGN} \in [0.1,1] \, \rm \micron$ range, and 18 sources with AGN fractions $>$ 0.5 with the $\lambda_{\rm AGN} \in [3,30] \, \rm \micron$ range, with 5 overlapping sources between these two populations. Altogether, from our $z>4$ sources with detections in IRAC ch3 and ch4, 27.8\% of them are potential AGN candidates. 

As seen in Fig. \ref{fig:color_classicAGN}, the majority of $z>4$ AGN candidates have $\beta_{\rm red}$ values greater than $-2$, but lower than $-1$, which corresponds to a flat continuum in the rest-frame optical range, and span a wide range of $\beta_{\rm blue}$ values. There is no significant difference between the median $\beta_{\rm red}$ and $\beta_{\rm blue}$ values of the two CIGALE-identified AGN populations ($\lambda_{\rm AGN} \in [0.1,1] \, \rm \micron$ and $\lambda_{\rm AGN} \in [3,30] \, \rm \micron$).  

In Fig. \ref{fig:color_classicAGN} we also include  AGN-classified sources from the literature. From \cite{Khostovan_2025}'s catalogue and the MIPS catalogue \citep{Sanders_2003}, we have 4 matching AGN sources above $z = 4$ each, though none of them are within the "Cut 1" or "Cut 2" colour selection. The \textit{Chandra} \citep{Civano_2016} catalogue yields 5 matching AGN for our whole $z>4$ source sample, and we have 8 matching sources with the AGN candidates from \cite{Andika_2025}. From the latter two catalogues, 20\% and 50\% of the \textit{Chandra} and \cite{Andika_2025} AGN candidates respectively are within our V-shaped-SED selection, as discussed previously.  

Overall, we see that 76\% (16 out of 21) of these classical literature AGN do not fall within our $\beta$-cut criteria. The X-ray detected AGN are mostly within the $\beta_{\rm red}$ cuts, but have slopes that are too positive in the rest-frame UV range, i.e. $\beta_{\rm blue}>-0.37$. Also, the $24 \, \rm \micron$-detected sources, which at $z>4$ should all correspond to AGN given the relatively shallow \textit{Spitzer} depths,  are sufficiently red in the rest-frame optical range, but do not make it in the $\beta_{\rm blue}$ cuts.  Finally, the spectroscopically identified AGN in the catalogue by \cite{Khostovan_2025} appear to fulfill the $\beta_{\rm blue}$ criteria, but they are not red enough to satisfy the $\beta_{\rm red}$ cuts. In summary, many of the  classical AGN have continuously increasing fluxes in the near-IR to mid-IR range, which excludes them from a V-shaped, double-power-law selection. These results are in line with the findings of \cite{Hainline_2025}.

Finally, as mentioned in the previous section, the \cite{Andika_2025} AGN candidates have been selected by SED fitting with CIGALE, but based on JWST data. In addition, the AGN classification criterion is an AGN fraction $\geq$ 0.2 in CIGALE's SED decomposition, which is more lenient than our criterion ($\geq 0.5$). In the end, not surprisingly the \cite{Andika_2025} AGN cover a wider range of $\beta_{\rm red}$ and $\beta_{\rm blue}$ values than our $z>4$ AGN.

\section{The \Euclid LRDs/LBDs in the context of the general LRD luminosity function at $z>4$} \label{sec:lrd_candidates}

\begin{figure}
    \centering
    \includegraphics[width=\columnwidth]{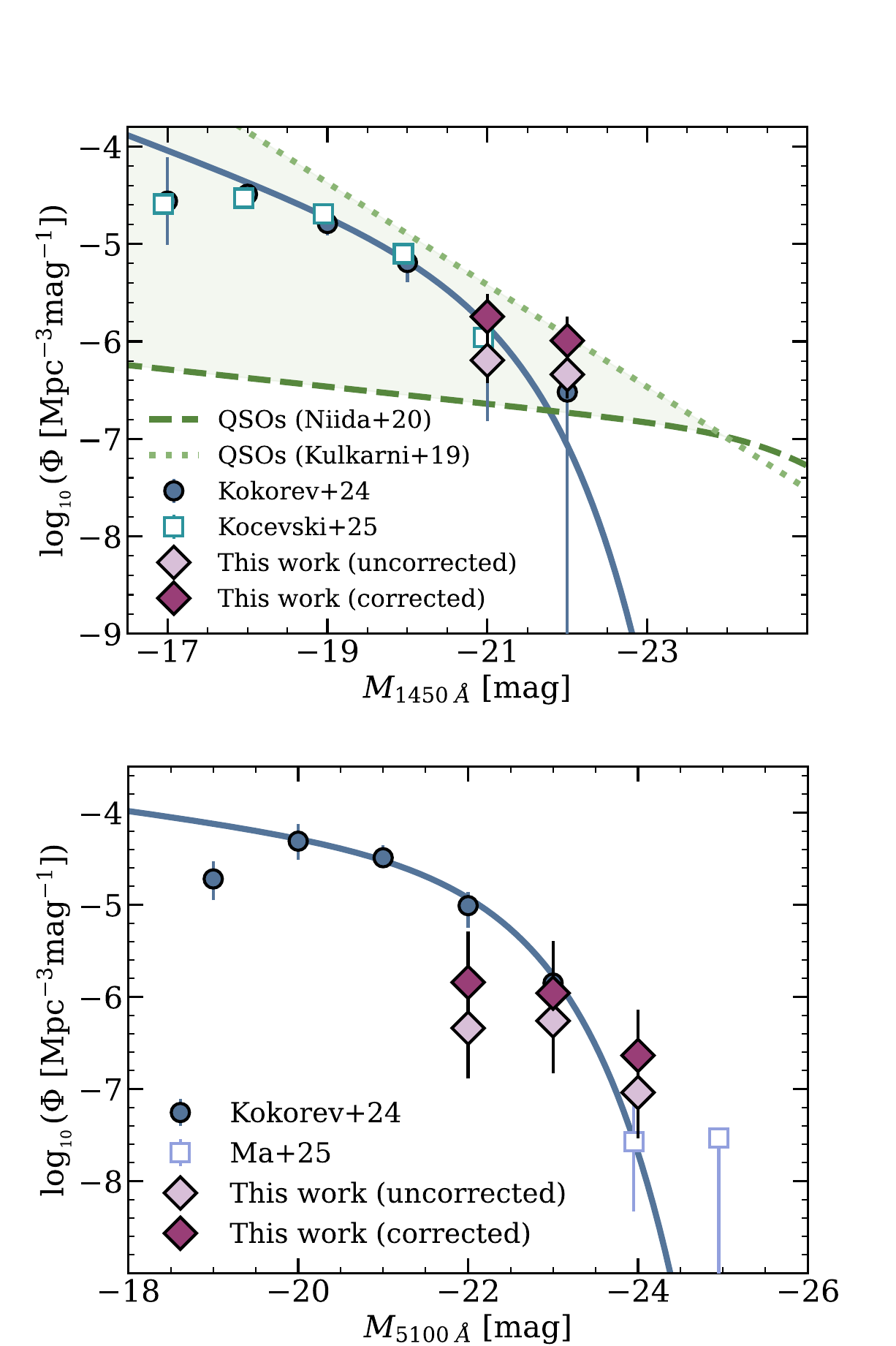}
    \caption{LRD/LBD luminosity function in the rest-frame UV ($\lambda_{\rm rest} = 1450 \, \text{\AA}$; \textit{top}) and optical ($\lambda_{\rm rest} = 5100 \,\text{\AA}$; \textit{bottom}), at $z  \in [4.5,6.5]$. The number densities are comoving. }
    \label{fig:lum_func}
\end{figure} 

Here our goal is to place the final sample of 16 \Euclid LRDs/LBDs in the context of the general JWST-selected-LRD luminosity function (LF) at $z>4$. We computed the rest-frame UV and rest-frame optical luminosity function of the \Euclid LRDs/LBDs using 
the standard $V_{\rm max}$ method \citep{Schmidt_1968}, after binning our sources in absolute magnitudes: in the rest-frame UV ($M_{1450 \, \text{\AA}}$) and optical ($M_{5100 \, \text{\AA}}$), as:

\begin{equation}
\Phi(x) = \frac{1}{\Delta x} \sum_i V_{\rm max}^{-1}(A,z_{\rm min},z_{\rm max}) \; ,
\end{equation}

\noindent where $\Delta x$ is the width of the $M_{1450 \, \text{\AA}}$ and $M_{5100 \, \text{\AA}}$ bins, and $V_{\rm max}$ is the maximum volume spanned by our \Euclid LRDs/LBDs. The maximum volume $V_{\rm max}$ is dependent on the survey area ($A$), and  the maximum redshift $z_{\rm max}$ at which a given source is still detectable within the  ($2\sigma$) detection limit of the galaxy survey. We calculated two versions of the luminosity function: one without any completeness or $V_{\rm max}$ corrections and another one with all the corrections (i.e., $V_{\rm max}$ and completeness corrections) incorporated, so that we can see how much the corrections affect the luminosity function. To correct for completeness we consider that $10\sigma$ detections correspond to 100\% completeness, and extrapolate based on the S/N ratios  of our sources.  

We show the resulting luminosity function in Fig. \ref{fig:lum_func}, along with results for JWST-selected LRDs \citep{Kokorev_2024,Kocevski_2024} and a very recent wide-area LRD identification \citep{Ma_2025}.  We fitted a Schechter function to our corrected LF number densities, using the equation

\begin{equation}
\rm \Phi(\it M)\rm \, d \it M \rm=(0.4\,ln\,10)\; \Phi^\ast \; 10^{0.4(\alpha +1)(\it{M^\ast} - M)} 
\rm  \, exp(-10^{0.4(\it{M^\ast} - M)}) \,d\it M\; .
\end{equation}

\noindent  We included the datapoints by \cite{Kokorev_2024} to constrain the fainter end. For the optical LF, we also included the datapoint by \cite{Ma_2025} in our fitting. The resulting Schechter best-fit parameters are shown in Table \ref{table:schechter_fit}.

We note that the comparison and joint fitting of our LRD/LBD LF data points with those of  \cite{Kokorev_2024} and \cite{Ma_2025} is meaningful. The criteria for JWST LRD selection has varied in the literature over the past years, but the SED slope cuts imposed by  \cite{Kokorev_2024} are very similar to those that we consider here.

\begin{table}
\begin{minipage}{\columnwidth}
  \centering
\scriptsize
\caption{Schechter function parameters for the LRD/LBD LF at $z \in [4.5,6.5]$. The number densities are comoving.} 
\label{table:schechter_fit}
\resizebox{\columnwidth}{!}{%
\begin{tabular}{p{1.9 cm}rr}
\toprule
Parameter & $ M_{1450 \rm \, \text{\AA}}$ & $ M_{5100 \rm \, \text{\AA}} $ \\ 
\midrule

$M^\ast$ [mag]& $-22.7\pm1.7$& $-22.0 \pm 0.29$\\ 
$\Phi^\ast $ $[\rm Mpc^{-3}]$ & $(0.95^{+1.95}_{-0.95})\times10^{-6}$& $(3.95\pm1.98)\times 10^{-5}$\\ 
$\alpha$ & $-1.83\pm0.19$& $-1.30\pm0.30$\\ 

    \hline  
\end{tabular}}

\end{minipage}
\end{table}

From these LF we see that our \Euclid-based data points cover the region around $M^\ast$ and even constrain the bright end of the optical LF at absolute magnitudes $M_{5100} \approx -24$ better than the \cite{Ma_2025} data. Interestingly, our results also show that the corresponding number densities of \Euclid-selected LRDs/LBDs are closer to those of classical QSOs at similar absolute magnitudes and redshifts. This is in contrast with the JWST-selected LRDs, whose number densities are about two orders of magnitude higher than similarly luminous QSOs \citep[Fig.~\ref{fig:lum_func}; see also][]{Kokorev_2024}. A likely reason for this difference is that our LRD/LBD sample is incomplete at its faintest magnitudes. The incompleteness could be driven by the \Euclid data photometric limits and also partly by our strict compactness criterion.  However, this could also raise the possibility that the \Euclid LRDs/LBDs may not necessarily have exactly the same nature as the JWST-selected LRDs/LBDs. Instead, being closer in number density to standard QSOs, our LRDs /LBDs could also be closer in nature to them. The spectroscopic follow up of our LRD/LBD candidates is necessary to test this scenario.

\section{Implications of the LRD selection technique} 
\label{sec:disc}

\subsection{Differences in the JWST and \Euclid LRD/LBD selection}
\label{sec:diff}

Here we applied a V-shaped SED technique to select our sources, which is similar to the technique used to select LRDs/LBDs with JWST data \citep{Kocevski_2024, Brazzini2026}.  There are, however, two main differences: i) the \Euclid data, even in the COSMOS field, is significantly shallower than typical JWST data at similar wavelengths, and therefore we can only select brighter analogues; ii) we made use of the $\HE$-band photometry as the bluest data point to calculate $\beta_{\rm red}$, which refers to shorter wavelength than the corresponding filter in the JWST data, namely F277W. In spite of being available in the COSMOS field, we explicitly did not use ancillary (ground-based) $K_{\rm s}$-band data for this purpose because it is in general not available in other \Euclid fields, so this source-selection technique cannot be replicated in all fields.

The main consequence of measuring the $\beta_{\rm red}$ slope on a longer baseline is the selection of a wider variety of objects. To fully appreciate the diversity of the selected objects we study the basic properties of the whole sample independently of compactness. In addition, our use of the $\HE$ band instead of a longer-wavelength filter results in the inclusion of sources with a red rest-frame optical continuum which would not be selected using the JWST F277W because of the presence of prominent emission lines in it (typically [\ion{O}{iii}]+H$\beta$ at $z \approx 4.5$). By selection, the latter sources are  missed in JWST LRD studies. 

In spite of these issues, we do not expect the nature of the LRDs/LBDs selected here to be substantially different to what they would be if we considered $K_{\rm s}$-band photometry. To test this, we repeated the sample selection process, using the $K_{\rm s}$ band as the bluest filter to measure $\beta_{\rm red}$. We found that the V-shaped SED sample using the $K_{\rm s}$ band was very similar to that using the $\HE$ band, with the $K_{\rm s}$-band selection able to reproduce 73\% of our V-shaped SED sources, and 75\% of our LRD candidates.  We note that the COSMOS $K_{\rm s}$-band photometry is mostly close to its depth limit for the V-shaped SED sources, which limits the number of sources selected using this filter. 

In particular for the COSMOS field, we can directly compare our V-shaped source selection with JWST-based LRD selections from the literature. \cite{Akins_2024} presented an `LRD' catalogue  within the COSMOS-Web area, albeit selected imposing only a colour cut in the rest-frame optical. Nonetheless, we searched for matching sources with our V-shaped-SED galaxies. Because the JWST data are deeper than our \Euclid-COSMOS data, we expect to select only the brighter end of their sources. Furthermore, because \cite{Akins_2024} imposes a much more stringent colour criterion on their LRD sources (F227W $-$ F444W $>$ 1.5), we expect to see few matches between their sources and our double power-law sources, even with the stricter "Cut 1" criteria. 

Indeed we find only 3 matches between our V-shaped SED sources and the \cite{Akins_2024}, with two of them having $\beta_{\rm red} >\, 0$ values, and the third has $\beta_{\rm red} = -0.03$, very close to the "Cut 1" $\beta_{\rm red}$  limit.  Interestingly, these 3 matched sources are not compact enough in the \Euclid \HE band to be included in our final LRD/LBD sample. They still have compactness greater than the median $z>4$ compactness value, but not higher than 1$\sigma$. This fact confirms that our compactness criterion based on the \Euclid \HE band is stricter than compactness defined in the JWST F277W.

\subsection{Using JWST LRD spectra as model templates for SED fitting}

Recent studies on LRDs have suggested that the blue continuum in their SEDs are not purely stellar in origin \citep[e.g.,][]{Inayoshi_Ho_2025}, and thereby may not be well fitted using standard galaxy templates \citep{Ronayne_2025}. To best describe LRD conditions, we would need model templates that include novel treatment of gas conditions, AGN and star-formation  \citep{Ronayne_2025}, which is currently beyond the scope of this work. Nonetheless, to explore the possibility of how using different model templates affects the output photometric redshifts and physical properties, we refit our V-shaped SED sources to a model template recently obtained from a stack of LRD spectra from \citep{Perez_Gonzalez_2026}. We checked which of our V-shaped SED sources preferred the LRD template as a solution (i.e. whether $\chi^2_{\rm LRD} \it < \chi^2_{\rm gal}$) and found that only 10\% did. Interestingly, we found a considerable overlap between the sources preferring LRD templates and the sources preferring QSO templates. About 36\% of the latter also prefer the LRD templates. However, given the choice between the LRD template and the QSO templates, all but one of these sources still prefer the QSO templates.

Out of out 16 robust \Euclid LRD/LBD candidates, we found that 25\% (4 out of 16) preferred the LRD model template.  Even though a significant percentage of our sources would be better fit with the LRD model template, we note that the reduced $\chi^2$ values are about unity in all cases, and with differences $<1$, which indicates that the galaxy, QSO, and LRD templates are similarly good for our sources and undistinguishable for all practical purposes, given the constraints provided by \Euclid and COSMOS ancillary photometry. 

\section{Comparison with BlueDOGs}\label{sec:bluedogs}

BlueDOGs are a rare class of DOGs, whose SED shapes are very similar to those of LRDs \citep{Noboriguchi_2023}. 
They are typically found at $z \in [2,3]$ and are significantly more luminous than our \Euclid-selected LRDs. The similarity between BlueDOGs and JWST LRDs is not just limited to their SEDs, though. The spectroscopic follow up of some BlueDOGs has revealed that they host BLAGN, as most LRDs do. Moreover, the BLAGN associated with BlueDOGs have extreme properties: they appear to accrete gas at several times the Eddington rate \citep{Noboriguchi_2022}.

\begin{figure}
    \centering
    \includegraphics[width=\columnwidth]{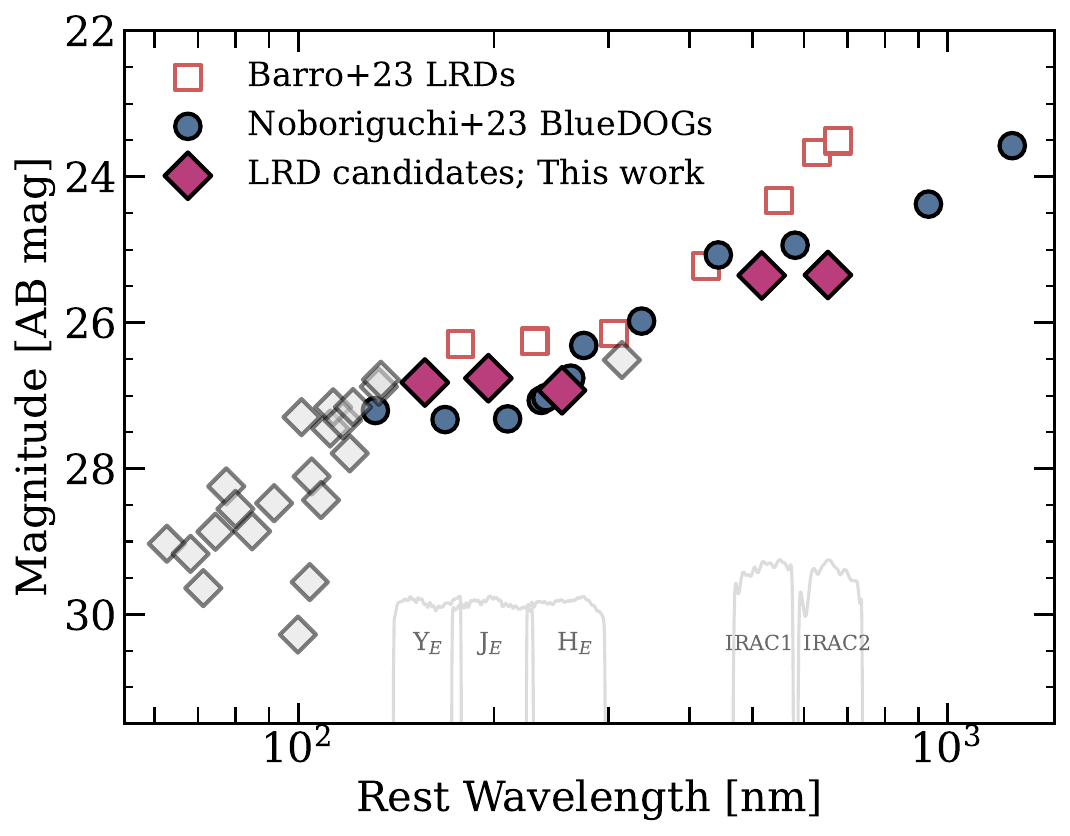}
    \caption{Average rest-frame SED of our final \Euclid-selected LRD candidates, compared with the average SEDs of JWST-selected LRDs and that of BlueDOGs. This figure is similar to the one presented in \citet{Noboriguchi_2023}, with this new version including our own data points for the \Euclid-selected LRDs.}
    \label{fig:bluedogs}
\end{figure}

The SEDs of our own \Euclid-selected LRDs/LBDs are on average also similar to those of JWST-selected LRDs and, in turn, also similar to the BlueDOG average SEDs, as can be seen in Fig.~\ref{fig:bluedogs}. However, probing the existence of BLAGN in the \Euclid-selected LRD/LBD candidates requires deep spectroscopic data, which unfortunately are currently not available in COSMOS for these kinds of sources. Future, dedicated spectroscopic follow-ups will reveal whether the \Euclid LRDs/LBDs do have a BLAGN. If this is the case, we will be able to estimate the corresponding black hole masses to locate our galaxies on the BH-mass versus host-galaxy stellar mass plane. This will allow us to fully understand whether our sources have a similar nature to the JWST LRDs, or could plausibly constitute a bridge between the JWST LRDs and classical QSOs.  

\section{Summary and conclusions}\label{sec:conclusion}

We have studied the properties of a population of \Euclid/IRAC-selected, V-shaped-SED galaxies at $z>4$ in $\sim 0.75 \, \rm deg^2$ of the COSMOS field. Our sample contains a total of 16 and 233 sources, for two different cuts in the rest-frame optical SED power-law slopes, namely $\beta_{\rm red}>0$ and $\beta_{\rm red}>-1$, respectively. Out of these sources, a total of 16 satisfy also having a compactness $1\sigma$ above the median compactness of all $z>4$ galaxies, as defined from their \Euclid $\HE$-band images. We consider these 16 sources to be the most robust \Euclid LRD/LBD candidates. We note that imposing compactness on the $\HE$ band is stricter than imposing it at longer wavelengths, as compactness typically increases with wavelength, so our selection is very conservative. 

The fact that only 2 out of 16 of our final LRD/LBD candidates satisfy the stricter  $\beta_{\rm red}>0$  criterion, which is that usually applied to JWST-selected LRDs,  indicates that: i) scaled-up versions of the reddest LRDs are rare; ii) relaxing the criterion to a less steep $\beta_{\rm red}$ slope provides a much wider sample of LBD candidates for spectroscopic follow up. This is in line with the fact the findings of \cite{Hainline_2025} and \cite{Brazzini2026} based on JWST data. 

The sources selected with  the most restrictive red SED slope  (i.e. $\beta_{\rm red}>0$)  are typically extended massive galaxies, with stellar masses $> \, 10^{10} \, \it M_\odot$, and low dust extinction, which are located around the star-formation MS of the $M_\ast$--SFR plane. This appears as a main difference with respect to the fainter JWST sources. In the brighter \Euclid sources at $z>4$, the host galaxy is generally detected and, therefore, most of them do not satisfy the compactness criterion imposed for LRD candidates. Instead, by relaxing the red slope cut to $\beta_{\rm red}>-1$, a more heterogeneous population, spanning about three dex in stellar mass, is selected.  

Based on their compactness, we identified a final sample of 16 scaled-up LRDs/LBDs. These objects are much more massive than JWST-selected LRDs and, interestingly, about a half of them appear to be as old as the Universe at their redshifts. This is in clear contrast with respect to the general population of sources with V-shaped SEDs, which span a wide range of ages from $\approx 0.01$ to $\approx 1.2 \, \rm Gyr$.

After cross-matching with  existing AGN catalogues and considering the SED-fitting results with \texttt{CIGALE}, we find that (at least) 9.4\% of the V-shaped-SED selected sources at $z>4$ could host an AGN. However, interestingly, most of the classically selected AGN via X-ray surveys and bright $24 \, \rm \micron$ sources at $z>4$ do not comply with the double power-law SED criteria, but rather constitute a disjoint population. This is consistent with the fact that LRDs are X-ray weak, and that in general most AGN do not have colours or $\beta$ values consistent with LRDs.

We have also placed the \Euclid LRDs/LBDs in the context of the general LRD LF at high redshifts. These sources constrain the LF around the knee, $M^\ast_\lambda$. The comoving number density of LRDs/LBDs at these absolute magnitudes is comparable to those of similarly luminous, standard QSOs. This is in contrast to what is observed at lower luminosities, for which the LRD number densities are up to two orders of magnitude higher. This suggests that the scaled-up LRDs/LBDs selected with \Euclid may be more similar to classical QSOs, although this is not apparent from their lack of X-ray and mid-infrared detections, so their properties could be intermediate. We hypothesize that some of these LRDs/LBDs could constitute a bridge between the JWST-selected LRDs and standard QSOs. In addition, the \Euclid-selected LRDs/LBDs are intermediate both in luminosity and number density between the JWST-selected LRDs and the very rare sources known as BlueDOGs \citep{Assef_2016, Noboriguchi_2019}. Some of these BlueDOGs have been spectroscopically confirmed to have LRD-like properties \citep{Noboriguchi_2022, Kim_2025}. For all these reasons, we conclude that a detailed spectroscopic follow up of the \Euclid LRDs/LBDs is utmost necessary to fully understand the nature of these sources and their role in the overall context of AGN and galaxy evolution.


\begin{acknowledgements}

AAT and YF acknowledge funding from the Dutch Research Council (NWO) through the award of an Open Competition ENW-XL Grant (P.I. Kuijken). KIC, RNC and GD acknowledge funding from the Dutch Research Council (NWO) through the award of the Vici Grant
VI.C.212.036 (P.I. Caputi). GG acknowledges the Kapteyn Astronomical Institute of the University of Groningen for hospitality while part of this work has been done.

\AckEC

\AckDRone
\cite{DR1cite}

\AckDatalabs


\end{acknowledgements}

%
\bibliographystyle{aa} 
\bibliography{sample701,Euclid} 
%

\begin{appendix}



\end{appendix}

\end{document}